\documentclass[Letter]{aa}
\usepackage[varg]{txfonts}
\usepackage{graphicx}
\usepackage{amssymb}
\usepackage{placeins}
\usepackage{float}

\usepackage{natbib,twoopt}
\usepackage[breaklinks=true]{hyperref}
\bibpunct{(}{)}{;}{a}{}{,}
  \newcommandtwoopt{\citeads}[3][][]{\href{http://adsabs.harvard.edu/abs/#3}%
    {\def\hyper@linkstart##1##2{}%
     \let\hyper@linkend\@empty\citealp[#1][#2]{#3}}}
  \newcommandtwoopt{\citepads}[3][][]{\href{http://adsabs.harvard.edu/abs/#3}
    {\def\hyper@linkstart##1##2{}%
     \let\hyper@linkend\@empty\citep[#1][#2]{#3}}}
  \newcommandtwoopt{\citetads}[3][][]{\href{http://adsabs.harvard.edu/abs/#3}%
    {\def\hyper@linkstart##1##2{}%
     \let\hyper@linkend\@empty\citet[#1][#2]{#3}}}
  \newcommandtwoopt{\citeyearads}[3][][]%
    {\href{http://adsabs.harvard.edu/abs/#3}
    {\def\hyper@linkstart##1##2{}%
     \let\hyper@linkend\@empty\citeyear[#1][#2]{#3}}}
\makeatother

\newcommand{\MC}{\multicolumn}
\newcommand{\kms}{km\,s$^{-1}$}

\newcommand{\HI}{H{\sc i}}
\newcommand{\HII}{\ion{H}{ii}}
\newcommand{\sunn}{$_{\odot}$}

\newcounter{qub}

\begin{document}

\title {The Peekaboo galaxy: New SALT spectroscopy and implications of archive HST data}

\author {A.Y.~Kniazev\inst{1,2,3}\thanks{E-mail: a.kniazev@saao.nrf.ac.za}
        \and
        S.A.~Pustilnik\inst{4}\thanks{E-mail: sap@sao.ru}}
\institute{%
South African Astronomical Observatory, PO Box 9, 7935 Observatory, Cape Town, South Africa
\and
Southern African Large Telescope Foundation, PO Box 9, 7935 Observatory, Cape Town, South Africa
\and
Sternberg Astronomical Institute, Lomonosov Moscow State University, Moscow, Russia
\and
Special Astrophysical Observatory of RAS, Nizhnij Arkhyz, Karachai-Circassia 369167, Russia}

\date{Accepted 2025 May 17, Received 2025 April 1}

\abstract
{The dwarf galaxy Peekaboo (HIPASSJ1131--31) was recently identified as a local volume (LV) gas-rich and
extremely metal-poor (XMP) dIrr. Its gas metallicity is Z$\sim$Z\sunn/50, with a $\pm$1$\sigma$ uncertainty range
of [Z\sunn/72--Z\sunn/35]). Its 'tip of the red-giant branch' distance is 6.8$\pm$0.7~Mpc. The Hubble
Space Telescope (HST) data for its individual stars
revealed that its older red-giant-branch stars comprise a smaller part of the galaxy, while the majority of
visible stars have ages of less than one
to a few gigayears. Thus, the Peekaboo dwarf can be considered as the nearest record-low Z dwarf.
As such, the galaxy deserves a deeper multi-method study that examines the properties of its young massive stars
and the fainter
older population as well as its ionised gas and the dominant baryonic component of \HI\ gas.}
{We aim to obtain the higher S-to-N SALT optical spectra of two \HII\ regions in Peekaboo in order to improve the accuracy
of its gas O/H and to determine abundances of Ne, S, N, and Ar.
With archive HST images, we aim to identify the hot massive stars, including exciting the two \HII\ regions
(i.e. east and west), and the XMP supergiants
as important targets for follow-up studies of their evolution with upcoming extremely large telescopes.}
{We used the direct (T$_{\rm e}$) method for the east \HII\ region in which a [O{\sc iii}]$\lambda$4363\AA\ line
is well detected in order to estimate its parameter 12+log(O/H).
In the west \HII\ region, the line [O{\sc iii}]$\lambda$4363\AA\ is not detected, so we estimated its O/H via the
empirical 'strong-line' method of Izotov et al. The resulting value of O/H is very close to that in the east \HII\ region.}
{The new spectroscopy of the Peekaboo dwarf allowed us to substantially improve the accuracy of its direct O/H estimate,
and we obtained 12+log(O/H) = 6.99$\pm$0.06~dex. The new data reveal that emission lines in the east region consist of two
components with a velocity difference of $\sim$65~\kms. The fainter approaching component could be related to a
fast-moving WR star thrown from a cluster or a binary system. Using the HST $V$ magnitudes and colour $V-I$, we identified
tentative O-type and very hot candidate WO stars, which are likely the ionising stars of the studied \HII\ regions.}
{With the new optical spectra, the Peekaboo galaxy is confirmed as the lowest-metallicity dwarf in the LV
and as a valuable object for in-depth multi-method studies.
We separate its most luminous stars for follow-up ground-based brightness monitoring and spectroscopy.
}

\maketitle
\nolinenumbers

\section[]{Introduction}
\label{sec:intro}

The 'local' gas-rich extremely metal-poor (XMP) dwarfs are interesting objects since they
challenge modern cosmological models of galaxy formation and evolution, and they
allow one to check the predictions of such models for the formation and survival of objects with extreme
properties. Based on the sum of their observational properties, gas-rich XMP dwarfs 
appear to be similar to galaxies in the early stages of evolution.

The existence of such unusual, very rare, so-called very young galaxies (VYGs) was predicted
in N-body simulations \citep{Tweed18}. The VYGs are galaxies that formed the major part of
their stellar mass within the last 1 Gyr. While VYGs are expected to be very rare, their
proportion in simulations strongly depends on
the type of dark matter employed (i.e. cold DM versus warm DM).

Apart from the limitations on the cosmological models that the statistics of XMP VYGs
can imply, their study is of interest in at least in two more areas.
Firstly, this is due to the expected similarity of their properties to those of young galaxies in the
early Universe (very gas rich and very metal poor). Being at the distances of
the Local Volume (LV) and its environs, XMP VYGs allow one to obtain much deeper insights
into their properties and to tie them to the processes of galaxy formation
in the early Universe.

The other important issue is related to stellar physics and galaxy
evolution. Namely, it concerns the basic understanding of the formation and evolution of
massive stars and their feedback into the interstellar medium of their parent galaxies \citep[e.g.][]{Eldridge22}.
For the cosmology of the early Universe, it is especially crucial to understand
the properties and the evolution of the massive stars with metallicities
tens to hundreds of times lower than the solar one Z\sunn.
Such massive, very low Z stars can be found in only a handful of XMP dwarf galaxies
within the LV and its environs \citep{Garcia21}. If identified, such XMP massive stars can be the
primary targets of spectroscopic studies using the next generation ground-based telescopes. In the meantime, the first steps
in this direction have already been presented for a nearby dIrr Leo~A with Z $\sim$ 0.03~Z\sunn\ by \citet{Gull22}
and for four low-metallicity galaxies, including Leo~P with Z $\sim$ 0.025~Z\sunn\ by \citet{Telford24}.

The Peekaboo galaxy, the optical counterpart of the \HI\ source HIPASS~J1131--31, got its name
from the fact that it was hidden within the halo of a nearby bright star \citep{Koribalski2018}.
The galaxy was recently described and studied by \citet{Peekaboo}.
As that paper reports, the "Peekaboo dwarf is a faint (M$_{\rm V}$ = --11.27~mag), the Local Volume
(LV) gas-rich dIrr at the TRGB -based distance of 6.8~Mpc (distance module = 29.2~mag).
Its gas metallicity, derived with the direct T$_{\rm e}$ method, appeared one of the lowest of
the known so far (12+log(O/H) = 6.99$\pm$0.16~dex)". The cited uncertainties of O/H
correspond to the $Z$ range of a factor of two: [Z\sunn/72--Z\sunn/35]). The authors also discussed
the probable "youth" of this galaxy and highlighted the abnormally small population of aged red-giant-branch stars.

In this letter, using new SALT spectroscopy, we update the estimate of Peekaboo's metallicity and
substantially improve on its
accuracy. We also used archive Hubble Space Telescope (HST) images and Extragalactic Distance Database
(EDD)\footnote{https://edd.ifa.hawaii.edu}
photometry of Peekaboo's stars to identify the hot massive stars ionising the two \HII\ regions
and several of the brightest supergiants. The latter can be attractive targets for HST and extremely large telescope
(ELT) spectral studies, as the record-low metallicity massive evolved stars.

The layout of the letter is as follows. In Sect.~\ref{txt:data_analysis}, the data and analysis are briefly described.
Sect.~\ref{txt:res} presents the main results.
These results are discussed in Sect.~\ref{txt:dis} and summarised in Sect.~\ref{txt:sum}. Details of the data and results are
shown in the appendices. The scale we adopted is 34~pc in 1~arcsec.

\begin{table*}
    \caption{Journal of the SALT observations.}
    \label{tab:log}
    \begin{tabular}{llccccc} \hline\\[-0.30cm]
        Date          & Grating & Exposure       & Spectral scale      &  Seeing   & Spectral range &  FWHM  \\
                      &         & (sec)          & (\AA\,pixel$^{-1}$) &  (arcsec) & (\AA)          &   (\AA)\\
        \hline\\[-0.30cm]
    2022 February 22$\dagger$ & PG900   & 1300$\times$2  &  0.97               &  1.60     & 3650$-$6700    &  4.7 \\
    2022 February 22$\dagger$ & PG1800  & 1300$\times$2  &  0.40               &  1.60     & 5950$-$7240    &  1.8 \\
    2022 February 26          & PG3000  & 1100$\times$2  &  0.24               &  1.80     & 3690$-$4440    &  1.1 \\
    2022 February 27          & PG3000  & 1100$\times$2  &  0.24               &  1.70     & 3690$-$4440    &  1.1 \\
    2022 March    09          & PG1800  & 1100$\times$2  &  0.40               &  1.50     & 5950$-$7240    &  1.8 \\
    2024 March    06          & PG3000  & 1100$\times$2  &  0.24               &  1.50     & 3690$-$4440    &  1.1 \\
    2024 March    07          & PG900   & 1100$\times$2  &  0.97               &  1.40     & 3650$-$6700    &  4.7 \\
\hline
\\[-0.35cm]
\multicolumn{7}{l}{$\dagger$ Spectra for this date were originally obtained for paper \citet{Peekaboo}.} \\
\hline
    \end{tabular}
\end{table*}

\section{Observations and data analysis}
\label{txt:data_analysis}

\subsection{SALT observations and analysis}
\label{ssec:SALT_data}

We carried out long-slit optical spectroscopic observations of the Peekaboo galaxy using the Robert Stobie Spectrograph
\citep[RSS;][]{Burgh03, Kobul03} installed at the Southern African Large Telescope \citep[SALT;][]{Buck06, Dono06}.
Different Volume Phase Holographic (VPH) gratings were used with the long slit of 1.5\arcsec\ by 8\arcmin\ to cover the
full spectral range from 3600~\AA\ to 7250~\AA. The position angle for all observations was 87.88\degr, and the slit position was exactly the same as shown in Figure~6 of \citet{Peekaboo}, where a reference star was used to put the slit in exactly the same position for different observations. A complete list of details is shown in Table~\ref{tab:log}, where all dates, seeing, covered spectral range, spectral scale, and full width at half maximum (FWHM) are shown.
The final spectrum taken with grating PG0900 was used as a basic one since it covers the greatest spectral range.
Spectral data obtained using other gratings were processed independently, and emission lines were measured in
the same spatial areas. After that, the line fluxes, measured on the spectra obtained with other gratings, were
recalculated to the flux system obtained using the PG900 grating. Namely, the total fluxes of the bright lines, measured on spectra for both gratings, were used to determine the conversion factor, and then the fluxes for the weaker lines were re-calculated using this factor. Spectrophotometric standards were observed at twilight as part of the SALT standard calibration plan.

Given that SALT is equipped with an atmospheric dispersion compensator (ADC), the effect of atmospheric dispersion
at an arbitrary long slit $PA$ is negligible.
Additionally, SALT's design incorporates the movement of the telescope's
pupil during tracking, thereby resulting in constant alteration of the telescope's effective area.
Therefore, accurate absolute photometry and spectrophotometry are not possible. However, relative flux calibration can
be used. Thus, the relative energy distribution in the spectra can be obtained with SALT data.

Fortunately in our research, we had the reference star on the slit and its known magnitudes in several bands from
the Legacy DR10 database \citet{Legacy}. Therefore, we used this star as a local standard to derive the line fluxes and luminosity of H$\beta$. (See details in Appendix, Sect.~\ref{sec:2dspectra}.)

The primary data reduction was done with the SALT science pipeline \citep{Cra2010}. The long-slit reduction was
done with the RSS pipeline described by \citet{Kniazev2022}. The 1D spectra extraction, emission lines measurement,
and calculation of chemical abundances was done in the way described by \citet{Sgr}. The fully reduced 2D spectra
of Peekaboo are shown in Figure~\ref{fig:SALT_2d_spec}, and the most prominent lines are labelled. Since the RSS detector is a mosaic and consists of three CCDs, the positions of two gaps between these CCDs are also shown.

Errors in line intensities have been propagated in the calculations of the reddening correction, electron temperatures,
and densities and have been propagated to the elemental abundance errors. The observed emission line intensities,
$\rm F(\lambda)$, relative to the F(H$\beta$) line and the intensities, $\rm I(\lambda)$, corrected for interstellar
extinction and underlying stellar absorption are presented in Table~\ref{t:Intens}. The equivalent width, $EW$(H$\beta$),
of the H$\beta$ line; the calculated absorption equivalent widths, $EW$(abs), of the Balmer lines; the extinction
coefficient, $C$(H$\beta$) -- which is a sum of the internal extinction in Peekaboo and foreground extinction in
the Milky Way; and the subsequently recalculated $E(B-V)$ value are also listed there.

\subsection{Hubble Space Telescope data}
\label{ssec: HST_data}

Hubble Space Telescope data are well described by  \citet{Peekaboo}.
Here, we used the information on positions, magnitudes, and colours of the hottest stars as the potential
exciting stars of the observed \HII\ regions. The evolved luminous stars (supergiants) are also identified
with these data, and they should be valuable in more advanced studies of the lowest metallicity massive stars
using upcoming facilities, including ELTs.

In Figure~\ref{fig:HST_color}, we show two zoomed-in HST images. In the top panel, we show an image in the F606W filter, and
all the stars are marked with their numbers in Table~\ref{tab:all-stars}. These are the same set of stars as in the paper by
\citet{Peekaboo}.
In the bottom panel, we present the colour HST image of the same area of Peekaboo taken from the HST
archive (Prog.~SNAP~15992, PI R.B.~Tully).

\section{Results}
\label{txt:res}

\subsection{Structure and kinematics of the main \HII\ regions }
\label{ssec:HII-regions}

Analysis of the long-slit spectra revealed the presence of two distinct \ion{H}{ii} regions.
Figure~\ref{fig:Ha_slit-profile} shows the profile of the H$\alpha$ line along the slit based on the spectral
data obtained with the PG900 grating. The figure clearly shows two regions,
which we refer to as regions `E' (east) and `W' (west) since the slit was oriented roughly in the east-west
direction across the galaxy. These regions appear as two peaks in the H$\alpha$ profile, although they are less
pronounced in the continuum profile.
Region W is fainter, with an extent of about 3~arcsec, while the brighter region E extends over about 5~arcsec.
We defined approximate boundaries for these regions (as shown in the figure) in order to extract 1D spectra along
the slit. The aim was to minimise spectral contamination between the regions while maximising the useful signal
from the emission lines. The selected region boundaries are also shown in the Figure~\ref{fig:SALT_2d_spec} as
red horizontal lines and in the top panel of Figure~\ref{fig:HST_color} as red circles centred on the reference
star with the corresponding radii. These boundaries were applied consistently to all the acquired spectral data.
The 1D spectra obtained by summing within these limits are shown in Figures~\ref{fig:SALT_1d_spec_900}
and \ref{fig:SALT_1d_spec_3000}.

A more detailed analysis of the extracted 1D spectra uncovered an interesting peculiarity. Namely,
all the observed emission lines in the spectrum of region E appear as a double component. Every line exhibits an asymmetric
blue wing. To test the hypothesis that all the lines represent two velocity components, we first fit each line with a single
Gaussian and then with a two-Gaussian model. The related improvement of the reduced $\chi^2$ statistic was
evaluated, and we found the improvement to be very significant. For various lines, it varies by a factor of $\sim$2 to 6.5.
Examples of the fitting for H$\beta$ and [\ion{O}{iii}]$\lambda5007$\AA\ lines and H$\delta$ and H$\gamma$,
for gratings PG900 and PG3000  are presented in
Figures~\ref{fig:SALT_1d_E_rb} and \ref{fig:SALT_1d_E_rb_PG3000}, respectively.

We calculated the radial velocities of the two components in the E region and that of the W region, taking the weighted
means on the three independent measurements with grisms PG900, PG1800, and PG3000. In turn, each of these velocities
was derived from the redshifts of six to seven of the strongest emission lines. The final values of V$_{\rm hel}$
for these three subsystems are as follows: East(red): 721$\pm$10~\kms; East(blue): 656$\pm$3~\kms; and West: 697$\pm$6~\kms.
For further discussion, we accepted these three Peekaboo ionised gas systems, of which East(red) and East(blue)
are co-spatial within the limits of our modest spatial resolution.

\subsection{New estimates of O/H and other elements}
\label{ssec:new_OH}

The measured and corrected line fluxes of atoms H and He and ions of O, N, Ne, S, and Ar and the derived element
abundances of O, N, Ne, S and Ar are presented in Tables~\ref{t:Intens} and \ref{t:Chem}.
Here, we briefly summarise the obtained results.

In Sect.~\ref{ssec:HII-regions}, we found that there are three separate subsystems on the slit.
For the East(red) subsystem, the line \mbox{[O\ {\sc iii}]$\lambda$4363} was detected at the level of 6.5$\sigma$.
This allowed us to directly determine the electron temperature, T$_{\rm e}$,  and derive the oxygen as well as other element
abundances via the
direct method. We obtained its 12+log(O/H)(T$_{\rm e}$) = 6.99$\pm$0.06~dex. The two non-direct methods, described in
Sect.~\ref{sec:lines},  give for the same subsystem, the O/H values within 0.02--0.03~dex of the former.
For the West and the East(blue) subsystems, the line [O\ {\sc iii}]$\lambda$4363 was not detected.
Their O/H were derived with the 'strong-line' method of \citet{Izotov19DR14}. For both objects, the derived O/H is within
0.02--0.03~dex from (O/H)(T$_{\rm e}$) of the East(red).

\subsection{Hubble Space Telescope data on XMP stars in the Peekaboo dwarf}
\label{ssec:HST_stars}

In the EDD \citep{Anand21} there is a table with parameters of all 56 individual stars within the Peekaboo
dwarf body. We sorted the stars by colour $(V-I)_{\mathrm 0}$ and separated the hottest stars
with $(V-I)_{\mathrm 0} \lesssim$ --0.25 (see Sect.~\ref{ssec:massive_stars}), which are expected to ionise
the surrounding gas.
We then looked for the stars that fall within (or near) the borders of the E and W \HII\ regions.
They are the stars most likely exciting these \HII\ regions.
 
There are also several supergiants in this XMP dwarf that are very interesting since they are the nearest
massive evolved stars with a record-low metallicity. They represent attractive targets for spectral study with HST
and next generation ELTs. (See the top panel of Fig.~\ref{fig:HST_color} with the HST image of Peekaboo, where the positions
of these stars are indicated.)
Most of the remaining stars with $(V-I)$ of $\sim$0.0--0.5~mag and M$_{\rm V} \lesssim$ --2.5~mag probably
belong to the blue helium burning type (BHeB) with ages less than $\sim$200~Myr \citep{McQuinn12}.

\section{Discussion}
\label{txt:dis}

\subsection{Metallicity of Peekaboo in context}
\label{ssec:OH_context}

The new determination of O/H in the Peekaboo dwarf is close to the earlier value but
has a much improved accuracy: 12+log(O/H)$_{\rm dir}$ = 6.99 $\pm$ 0.06~dex. We compared this value with that of
known dwarfs that have the lowest metallicities in the LV. To the best of our knowledge, within the LV
(R$\lesssim$11 Mpc), apart from Peekaboo, ten XMP dwarfs with 12+log(O/H) $\leq$ 7.17~dex
(or Z(gas) $\leq$ 0.03 Z\sunn) have been found. Six of them are published in \citet{J0926,BTA1,SALT2,Skillman13,ITG12}.
(See Table~\ref{tab:XMP-LV} in Appendix, where we compile their properties.)
Five of these six dwarfs reside within the nearby voids from \citet{PTM19}.
In the framework of the ongoing project 'Studying the void galaxies in the Local Volume', we found four new
LV XMP dwarfs, currently prepared for publication.
Only four of the total 11 LV XMP galaxies (including Peekaboo) have their O/H derived with the direct method.

The dwarf Leo~P, with 12+log(O/H) =7.17~dex \citep{Skillman13}, is the nearest known XMP galaxy, at $\sim$1.6~Mpc from the
Local Group centre. Its proximity allows for study of its massive hot XMP star O7-8~V with ground-based spectroscopy
and important constraints of its wind properties to be obtained \citep{Telford23, Telford24}. The remaining
LV XMP dwarfs are situated much farther away, at the distances of $\sim$7--11~Mpc. Thus, spectroscopy of their luminous
stars should be available with HST and upcoming ELTs.

\subsection{Hot and luminous stars, and blue outliers}
\label{ssec:massive_stars}

The proximity of Peekaboo and the high resolution of the HST images allow one to identify the hottest massive stars,
including stars ionising the E and W \HII\ regions.
We separate in the diagram of M$_{\rm V}$ versus $(V-I)$
in Fig.~\ref{fig:cmd} a region where all O-stars (for luminosity classes from V to I) fall according to the state-of-the-art
models at a metallicity of Z = Z\sunn/10 presented by \citet{Lorenzo25}. This area is indicated by a rectangle
with boundaries of
$(V-I)_{\mathrm 0}$ from --0.32 to --0.25~mag and M$_{\rm V}$ between --3.30 and --6.15~mag.
A list of the stars in this area and their additional parameters is presented in the appendix.
In Fig.~\ref{fig:cmd}, we also identify four supergiant stars, the evolved massive stars, conditionally with
M$_{\rm V} \leq $ --6.0~mag (or 21.4 $\leq$ $V \lesssim$ 23.5~mag). They probably represent the upper part of blue helium
burning stars, according to \citet{McQuinn12}.
They are interesting for detailed studies, as they represent the nearest evolved massive stars with
a record-low metallicity.

In particular, they can be observed spectroscopically at superb conditions (i.e. narrow slit and a seeing
of $\leq$0.5~arcsec)
with the Very Large Telescope at ESO. In addition, there is a good opportunity to check their possible large-amplitude
optical variability via monitoring of the entire galaxy and to analyse the light variations of the small regions around
the supergiants.
A successful example of monitoring the variability of the XMP LBV and other supergiants in \HII\ regions of the XMP galaxy
DDO68 \citep[12+log(O/H) = 6.98 -- 7.3~dex,][]{DDO68,IT07,Annibali2019} was recently presented in \citet{DDO68NR, DDO68-V1}.

Another type of evolved massive star can show up in a $V$ versus $(V-I)$ diagram as a 'blue outlier'
\citep[e.g.][]{Lorenzo22}.
Such a star has M$_{\rm V} \sim$ --3.5 to --5~mag and colours bluer than the blue edge of the Main Sequence stars (namely,
$(V-I) \sim$ \mbox{--0.32}; \citet{Lorenzo25}), and they can be helium burning WO stars with a surface temperature
of up to 150--200 kK \citep[][and references therein]{Lorenzo22}.

\subsection{Properties of Peekaboo's \HII\ regions}
\label{ssec:sizes_HII}

In section~\ref{sec:2dspectra} of the appendix, we present the distribution of H$\alpha$ flux along the slit, which
looks similar to two close Gaussians with FWHMs of $\sim$3.4 and 3.7~arcsec, respectively. Their sizes along
the slit (corrected for a seeing of $\sim$1.5~arcsec) are FWHM $\sim$ 3.05 and 3.4~arcsec, or $\sim$116~pc and 128 pc.
In the colour HST image,
one does not see such clear extended structures. These sizes probably represent the extent of the (smoothed due to seeing)
collection of several smaller \HII\ regions elongated along the galaxy major axis. Several of the smaller brightest
\HII\ regions are marked in Fig.~\ref{fig:HST_color}, and their parameters are summarised in Table~\ref{tab:all-HII}.

As the two-component velocity profiles of all emission lines in the E \HII\ region evidence, this cell of ionised
gas indeed has a complex origin. Roughly 30\% of its nebular emission is related to the ionised gas,
moving
with a radial velocity of about --60~\kms\ ('blue' component hereafter) relative to the main body, which has a system
velocity of V$_{\rm {helio, \HI}} =$ 716~\kms\  \citep{Peekaboo}.
The radial velocities of the E region's more massive 'red' velocity component and that of the
W \HII\ region are close to the system one.

The ionised gas of the 'blue' component should be connected with a fast-moving hot star. Such massive stars, which are
products of
co-evolution in binary systems, are well known in the Galaxy. Due to the 'poor' spatial resolution
of the long-slit spectra, the size of the 'blue' component cannot be addressed directly. However, since the 'blue'
component gas moves together with its exciting massive star, one can suggest that this shell is related to the mass loss
from the fast-moving star that occurred after the epoch when it was thrown out from a close binary (or a cluster).
 The abundance pattern in E(blue), similar to those in the other regions, indicates that its nebular emission originates
in the swept-up ambient gas shell. The case of the complex nature of the E \HII\ region partly resembles the case of
the \HII\ region A2 in IC4662. In this region,
the echelle spectroscopy revealed three components, and one of them is related to a known WR star and moves with a relative
velocity of $\delta$V $\sim$ 20~\kms\ \citep{Kniazev2025}.

In the appendix (Sect.~\ref{sec:2dspectra}), we estimate the luminosities of the line H$\beta$ of the three
emission-line subsystems.
Respectively, they are 7.5, 8.4, and 14.8$\times$10$^{36}$ erg~s$^{-1}$ for the W, E(blue), and E(red) subsystems.
These luminosities can be used to estimate the conditional number of hot stars ionising these regions. For further
calculations, we adopted the relation between L(H$\beta$) and the related flux of ionising quanta Q$_{\mathrm 0}$:
L(H$\beta$) = 4.76$\times$10$^{-13}$ Q$_{\mathrm 0}$. The resulting values of Q$_{\mathrm 0}$ for the three subsystems are as
follows: (1.6, 1.8,  and 3.2) 10$^{49}$ phot~s$^{-1}$. Recently, \citet{Telford23} estimated the value of Q$_{\mathrm 0}$
for the lowest accessible metallicity O7V star LP26 in Leo~P (12+log(O/H)=7.17~dex). This number,
(3.74$\pm$0.67)$\times$10$^{48}$ phot~s$^{-1}$, is $\sim$2.5 times lower than the value usually adopted for the O7V star with
Z = Z\sunn. That is, we need from  three to six O7V XMP stars per subsystem to provide the required amount of ionising
photons.

Comparison of the star positions (marked with their numbers) in the HST image  (top panel of Fig.~\ref{fig:HST_color})
and their positions in the diagram of Fig.~\ref{fig:cmd} revealed within the borders of the E region only one
certain luminous O star (No.~7) and about 1-mag fainter probable O stars (No.~5 and No.~6). Besides, two likely
luminous WR stars (No.~2 and No. 4) reside in the middle of the E region. One candidate very hot WR star, No.~1,
is close to No.~2. In particular, their extremely blue $V-I$ colours are consistent with those of WO stars, for which
their temperatures are estimated to be in the range 150-210 kK \citep[e.g.][]{Tramper2015}. Their parameter Q$_{\mathrm 0}$
falls in the range of (0.8--3)$\times$10$^{49}$ phot~s$^{-1}$. That is, even one such WO star is sufficient to provide
ionising photons for each of the three \HII\ subsystems.

Since observations indicate
that most WR stars are situated on the edges of the related \HII\ regions \citep[e.g.][]{Crowther2009}, this implies
that they were ejected from their parent cluster. Therefore, one can expect that the 'fast-moving' 'blue' subsystem
of the E region
is related to one of the above WR stars. The spectra of this region with a seeing of $\lesssim$0.5~arcsec
could help disentangle various subsystems within this region.

Apart from these two elongated E and W complexes, numerous small blue non-stellar objects ('nebulosities') are distributed
along the whole dwarf body. Their typical sizes of 0.2--0.5~arcsec correspond to linear sizes of 7--15 pc, or to
Str{\"o}mgren radii of $\sim$4--8~pc. Such small \HII\ regions are probably ionised and excited by individual hot
massive stars.
Their ($V-I$) colours are bluer than those of hot O-type stars.
Such very blue ($V-I$) colours were predicted for the nebular emission in low-metallicity high-temperature \HII\ regions in
\citet{Zackrisson01, Anders03}. In particular, \citet{Anders03} showed that the colour ($V-I$) for young starbursts with
Z$\sim$Z\sunn/50 can vary in the range of --0.7 to --0.25~mag for ages from 5 to 15~Myr due to the very strong
contribution of emission lines in the $V$-band. This is exactly the range we observed for the five regions N1--N5 shown in
Fig.~\ref{fig:cmd} and in Tab.~\ref{tab:all-HII}. As the lowest metallicity individual \HII\ regions in the nearby Universe,
these objects deserve spectroscopy with HST and/or ELTs.

\section{Conclusions}
\label{txt:sum}

The goals of this work was to improve the accuracy of Peekaboo's gas metallicity and to address properties
of its massive stars.
Based on the new data, we have presented on the dwarf galaxy Peekaboo and their discussion,
we arrived at the following conclusions:

\begin{enumerate}
\item
The new spectral data for the two \HII\ regions of the Peekaboo dwarf allowed us to uncover three ionised gas systems.
Two systems, E and W, are separated spatially by $\sim$4~arcsec between their centres (140~pc) approximately along the
galaxy major axis. In the E \HII\ region, we observed two 'co-spatial' kinematically separated gas systems with radial
velocities of V$_{\rm hel}$ of 720~\kms\ and 654~\kms ('red' and 'blue' components).
\item Thanks to the improved S-to-N ratio in the principal faint line [O{\sc iii}]$\lambda$4363\AA\ of the E region
(red), we derived its O/H with the direct (T$_{\rm e}$)
method: 12+log(O/H) = 6.99$\pm$0.06~dex. For the E(blue) subsystem and for the W \HII\ region,
the line [O{\sc iii}]$\lambda$4363\AA\ is undetected.
Their O/H values, derived via the empirical 'strong-line' method, appear
very close to that of the E(red) subsystem with (O/H)(T$_{\rm e}$).
The Peekaboo galaxy is the lowest metallicity dwarf within the LV and its environs among galaxies with gas O/H derived
via the direct method.
\item
We also derived the abundances of other heavy elements X (Ne, Ar, N, S), and they are consistent with the average X/O
 typical of galaxies in the lowest metallicity range.
\item
Using the EDD coordinates, magnitudes, and colours of individual stars in the body of the Peekaboo dwarf, we identified the
  hot massive stars (including candidate WO stars) tentatively ionising its \HII\ regions as well as four supergiants
  in this galaxy.
  These supergiants, as the brightest and the lowest-metallicity evolved massive stars, are valuable targets for 
  detailed studies with HST and the upcoming ELTs as well as for ground-based monitoring of their possible
 large-amplitude variability.
Their colours and spectral features as well as their variability could be suitable data to confront them with predictions
of the modern models of stellar  evolution at Z $\sim$ 0.02~Z\sunn.
\end{enumerate}

Thus, the Peekaboo dwarf is among the most intriguing dwarf galaxies in the Local Volume, and as such, it warrants
intensive, multi-method, in-depth study. Its properties -- ranging from high-resolution \HI\ gas morphology and kinematics,
through compact star-forming regions, to the lowest-metallicity known massive stars revealed via resolved space-based
photometry and spectroscopy -- are expected to significantly advance our understanding of the first building blocks
of the early Universe.

\begin{acknowledgements}

Spectral observations reported in this paper were obtained with the Southern
African Large Telescope (SALT) programs 2021-1-MLT-001 and 2023-1-MLT-006.
A.\,K. acknowledges support from the National Research Foundation (NRF) of South Africa.
The study was conducted under the state assignment of Lomonosov Moscow State University.
The work was also performed as part of the SAO RAS government contract by the
Ministry of Science and Higher Education of the Russian Federation. We acknowledge comments and
suggestions of the anonymous referee, which helped to improve the paper clarity.
This work is based on observations made with the NASA/ESA HST on the program SNAP~15992,
PI R.B.~Tully.
We acknowledge the use of the Extragalactic Distance Database and, in particular, its CMDs/TRGB part.
This work has made use of data from the Legacy Survey (DR10). The Legacy Survey team is
acknowledged for making their data products available to the scientific community.
\end{acknowledgements}

\begin{appendix}
\onecolumn

\section{Hubble Space Telescope image of the Peekaboo dwarf with the slit position}
\label{sec:slit}
\FloatBarrier

\begin{figure*}[!h]
    \centering{
	\includegraphics[clip=,angle=0,width=0.56\textwidth]{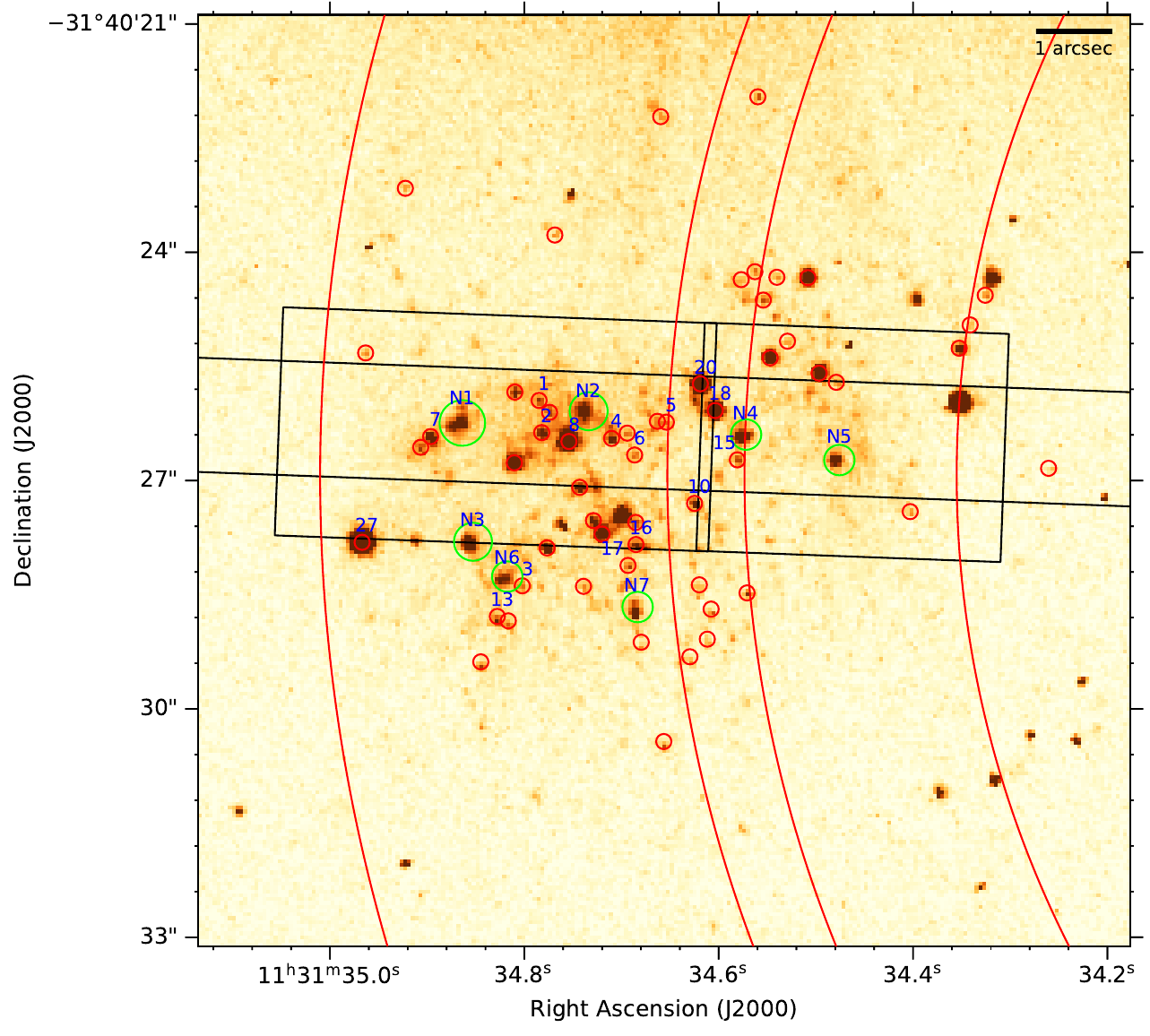}
\vspace{1.2cm}
	\includegraphics[clip=,angle=0,width=0.56\textwidth]{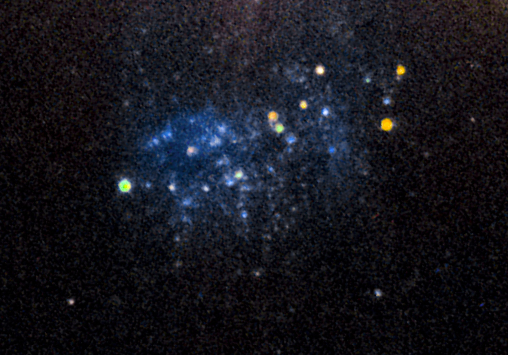}
    }
    \caption{  {\bf Top.}
Zoomed-in image of the Peekaboo dwarf in the F606W band from HST with the position of the
     1.5-arcsec-width long slit superimposed (similar to that shown in \citet{Peekaboo}). We expanded the boundaries
     of the slit width by 0.7~arcsec to account for the light of the adjacent region which falls to the slit due
     to the seeing effect.
     Positions of the discussed hot and supergiant stars (Sect.~\ref{sec:HST_stars}) are shown by the small circles
     with numbers, corresponding to the list in Table~\ref{tab:all-stars}.
     {\bf Bottom.}
     Colour ($V,I$ filters combined) image from HST of the Peekaboo
     dwarf with a similar scale and orientation as the top image. Numerous small (fraction of 1~arcsec) blue non-stellar
     objects ('nebulosities') are seen, with the linear
     sizes of 7--15~pc.  Due to the small sizes, they probably are related to singular massive stars. Seven of them
     are marked as N1--N7 in the top panel. Objects N1--N5 contribute to the emission in the slit. }
        \label{fig:HST_color}
\end{figure*}

In the top panel of Fig.~\ref{fig:HST_color} we show the zoomed-in F606W HST image of the Peekaboo dwarf.
The position of the 1.5-arcsec-width long slit is superimposed.
 We also overplot approximate boundaries of W and E \HII\ regions with arcs, defined through the distance
 from the 'reference' star.
The hottest and supergiant stars are marked to show their location relative to W and E \HII\ regions.

The bottom panel presents the same scale colour HST image of the galaxy. The brightest stars are
visible along with numerous blue non-stellar objects ('nebulosities') with the sizes of a fraction of arcsec.
The respective linear extent is  8 to 15 pc. These are likely small \HII\ regions, excited by individual
massive hot stars. They are marked as N1--N7 in the top panel. Probably, the integrated nebular emission collected on
the long slit, is just a sum of such smaller \HII\ regions, which, due to the smoothing effect of seeing, show up only as
emission of regions with sizes of $\lesssim$100~pc ($\lesssim$3~arcsec).

\clearpage
\section{Two-dimensional spectra, sizes of \HII\ regions, and related issues}
\label{sec:2dspectra}
\FloatBarrier

\begin{figure*}[!hb]
    \centering{
	\includegraphics[clip=,angle=0,width=0.80\textwidth]{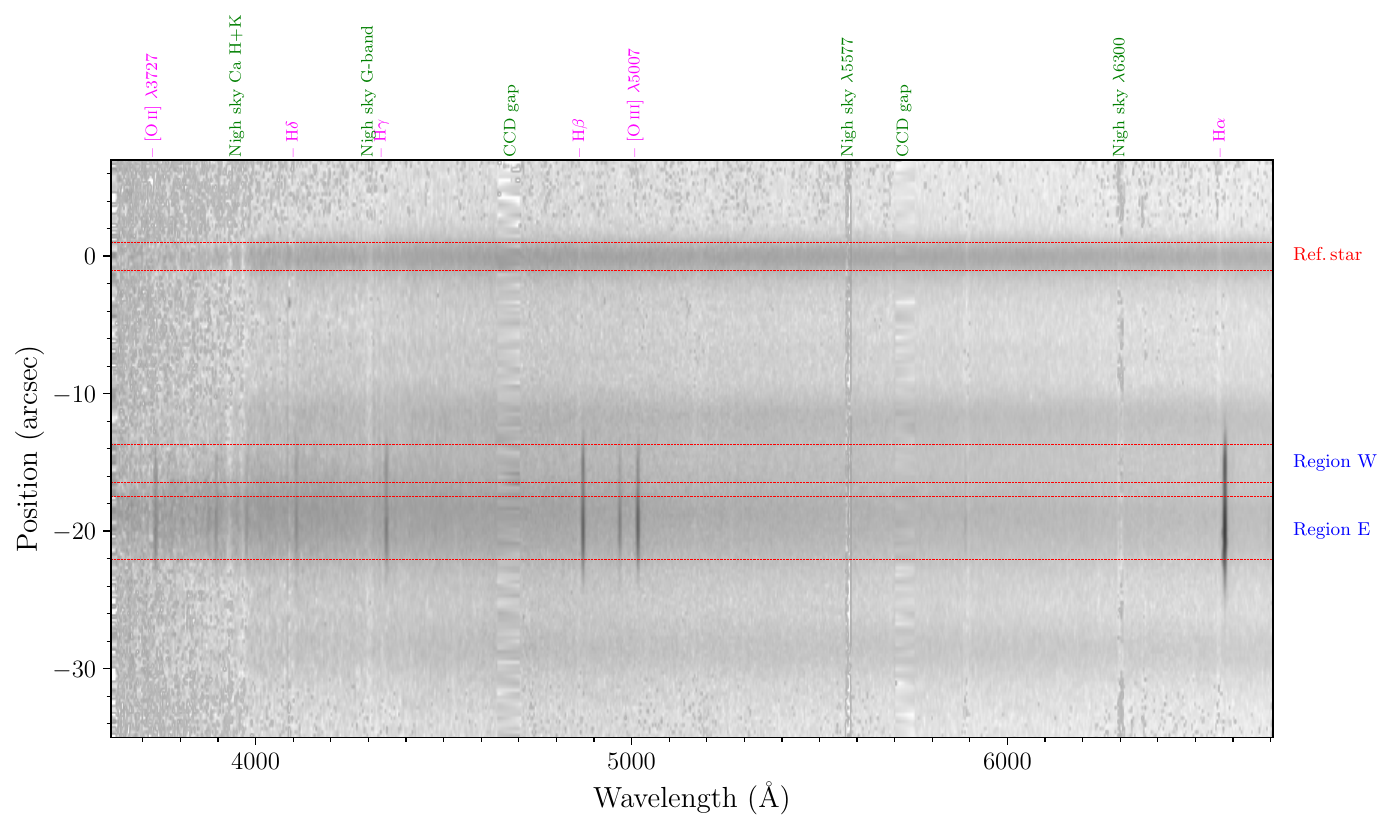}\\
	\includegraphics[clip=,angle=0,width=0.80\textwidth]{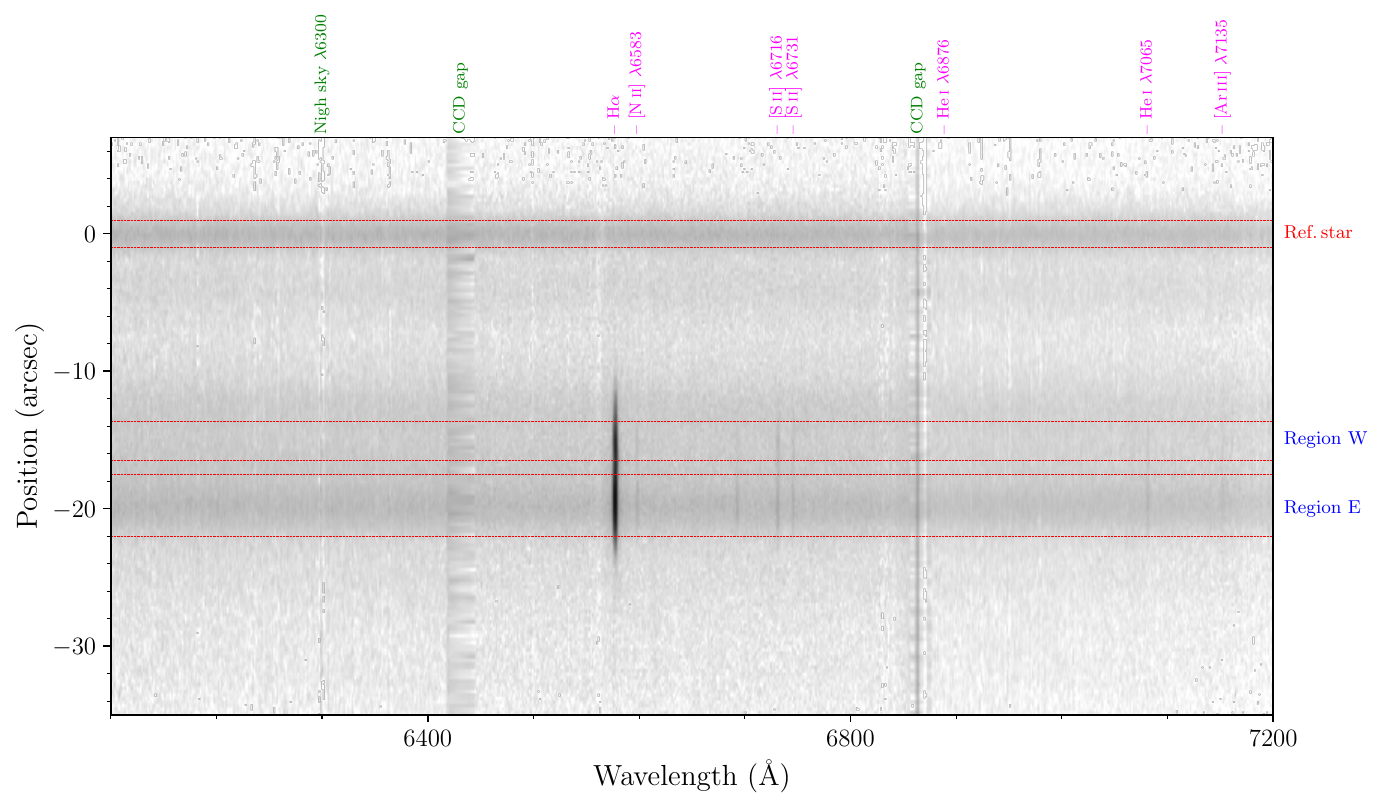}
    }
    \caption{Two-dimensional low-resolution spectra of the Peekaboo galaxy obtained with grating PG900 (top)
        and PG1800 (bottom). The logarithmic scale of intensities was used.
        Negative offsets are east of the reference star.
        The most prominent emission lines are indicated as well as CCD gaps and night sky lines. 
        \label{fig:SALT_2d_spec}}
\end{figure*}

In Fig.~\ref{fig:SALT_2d_spec} we present 2D spectra of Peekaboo with gratings PG900 and PG1800.
Positions of E and W \HII\ regions along the slit as well as that of the reference star are indicated.
The reference star (RA,Dec = 172.8887,--31.6742) has the available photometry in Legacy DR10 release \citep{Legacy}
with the following magnitudes: $g$=19.93, $r$=19.49, $i$=19.35, $z$=19.31.
We use this photometry and its 1D spectrum (we extract its full extent along the slit and estimate the loss on the
slit of 24\%) to produce the flux-calibrated spectra of the E and W regions and to estimate fluxes in H$\beta$ lines.
They, in turn, are used to estimate
luminosities in this lines and the related parameters of exciting hot stars in these regions.

The fluxes of emission H$\beta$ line of W and E regions are 8.5 and 26.3, respectively (in units of 10$^{-16}$
erg~cm$^{-2}$~s$^{-1}$). The line flux for the E region comprises two components, with the 'blue' component
part of 9.6 and the 'red' component of 16.7. The related line luminosities,  after correction for the measured
extinction in the \HII\ regions, C(H$\beta$)$\sim$0.18~dex, or factor of 1.6,
are as follows. For the W region, it is  7.5$\times$10$^{36}$, and for the E region, it is
23.2$\times$10$^{36}$ erg~s$^{-1}$, with the division
of (8.4 and 14.8)$\times$10$^{36}$ for the blue and red components, respectively.

\begin{figure}
    \centering{
        \includegraphics[clip=,angle=0,width=0.47\textwidth]{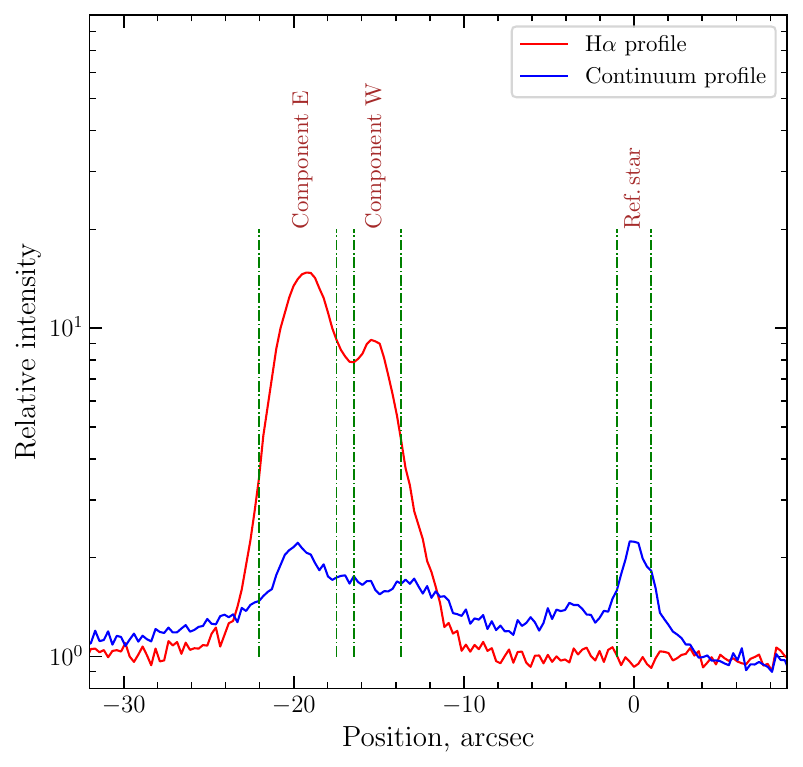}
    }
    \caption{Brightness distribution in the line H$\alpha$ along the slit
     and the cuts, adopted to extract individual spectra of the W and E regions. The estimated FWHM sizes
     are $\sim$2.0, 3.7 and 3.4~arcsec for the reference star, and for W and E components, respectively.
     After correcting for the seeing of 1.5~arcsec, their intrinsic FWHMs are $\sim$3.0 and 3.4~arcsec.
        \label{fig:Ha_slit-profile}}
\end{figure}

\clearpage
\section{Extracted one-dimensional spectra of the E and W \HII\ regions}
\label{sec:1d}
\FloatBarrier

\begin{figure*}[!h]
    \centering{
        \includegraphics[clip=,angle=0,width=0.70\textwidth]{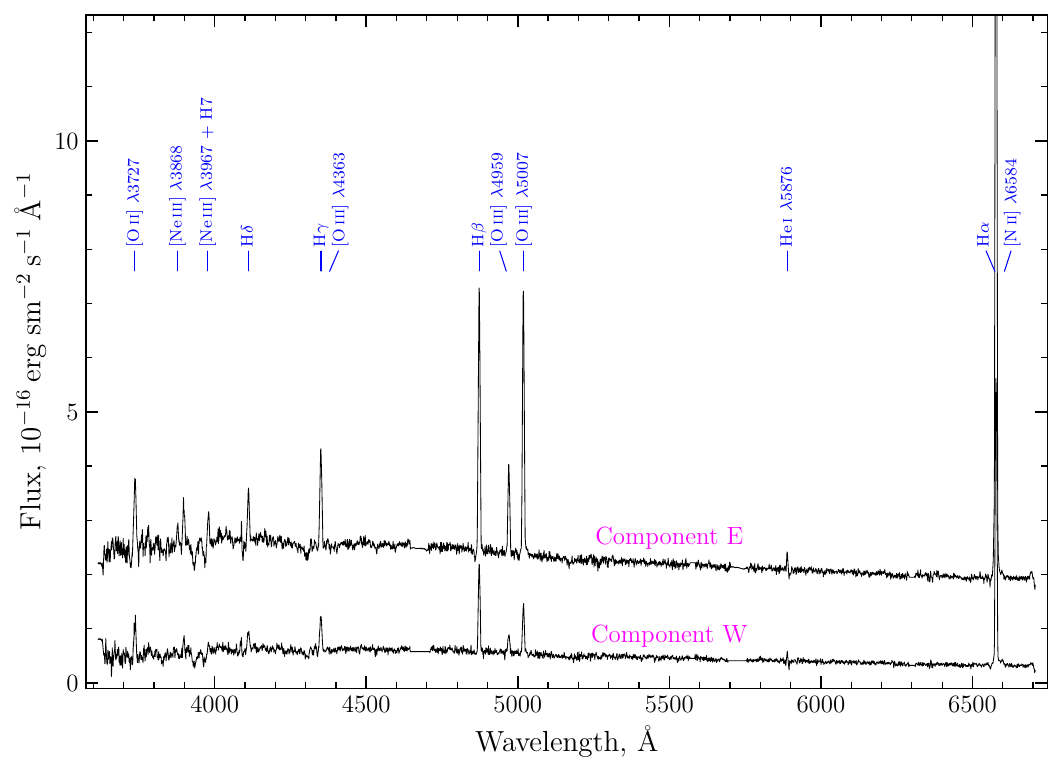}
    }
    \caption{Extracted 1D spectra of E and W \HII\ regions for grism PG900.
        Spectrum of region E is shifted up by 1.0 unit for the better view.
        The most prominent emission lines are indicated.
        \label{fig:SALT_1d_spec_900}}
\end{figure*}

\begin{figure*}[!h]
    \centering{
        \includegraphics[clip=,angle=0,width=0.70\textwidth]{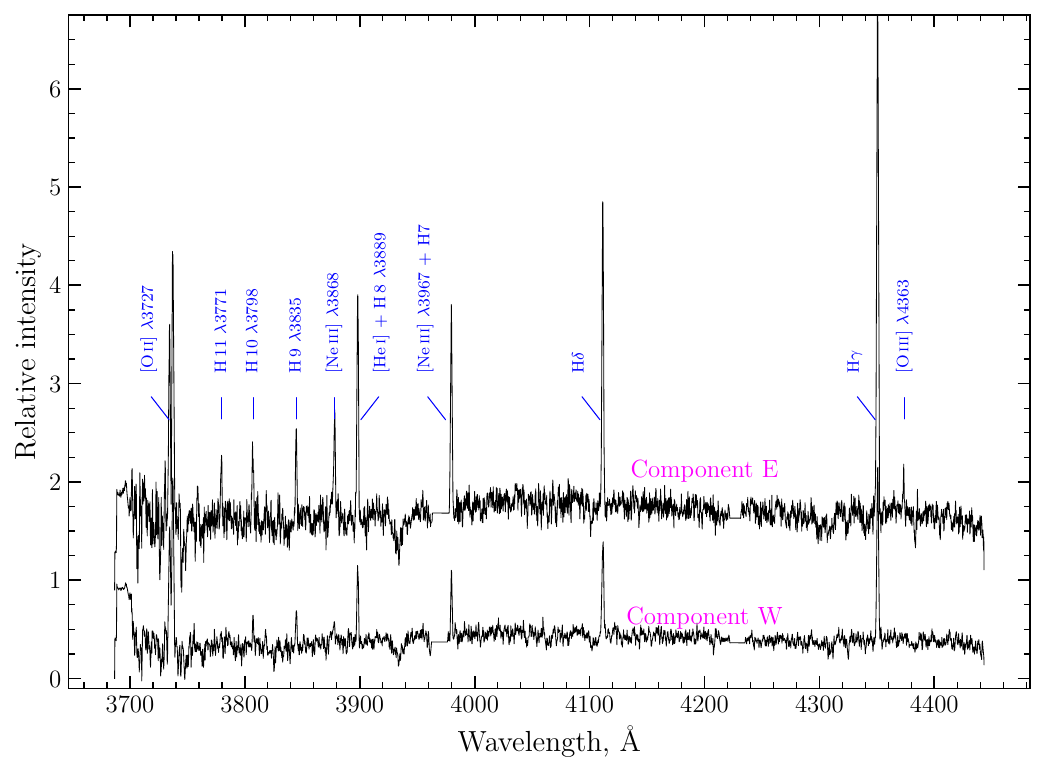}
    }
    \caption{Extracted 1D spectra of Peekaboo's E and W \HII\ regions for grism PG3000.
        Spectrum of region E is shifted by $\delta~y =$ +0.9  for the better view.
        The most prominent emission lines are indicated.
        \label{fig:SALT_1d_spec_3000}}
\end{figure*}

In Figure~\ref{fig:SALT_1d_spec_900} we show 1D spectra of Peekaboo's E and W regions with grism PG900,
in which we mark the main emission lines used for the abundance determination (Hydrogen Balmer lines
and lines of  O, Ne, N, S, Ar ions). The spectra are flux-calibrated via the known Legacy DR10
  photometry of the reference star.
In Figure~\ref{fig:SALT_1d_spec_3000}, a similar 1D spectrum is shown, with the higher spectral resolution
with the use of grism PG3000. The faint auroral line [O{\sc iii}]$\lambda$4363~\AA\ is well seen near H$\gamma$ in
the spectrum of the eastern \HII\ region.


\begin{figure}
    \centering{
        \includegraphics[clip=,angle=0,width=0.40\textwidth]{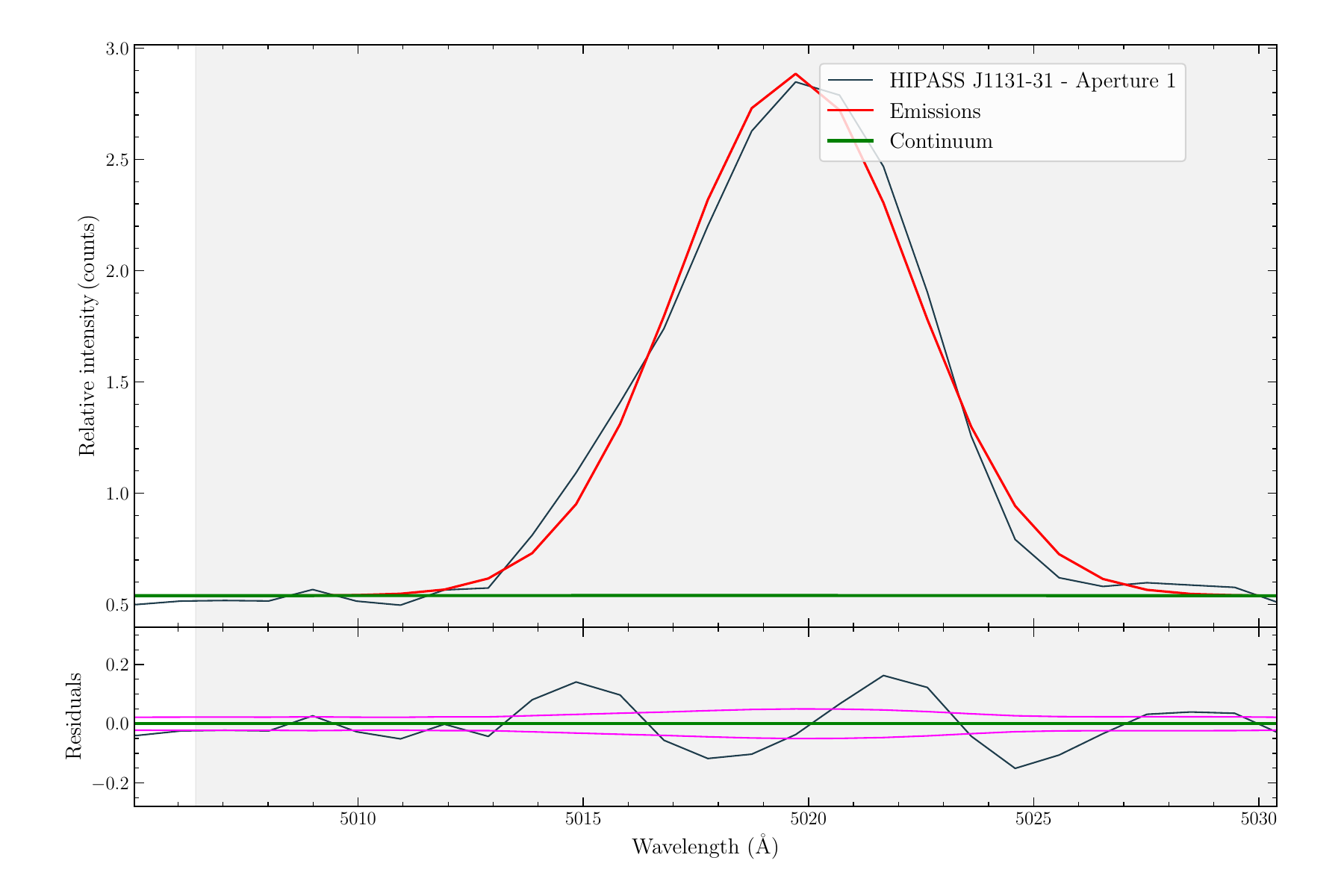}
        \includegraphics[clip=,angle=0,width=0.40\textwidth]{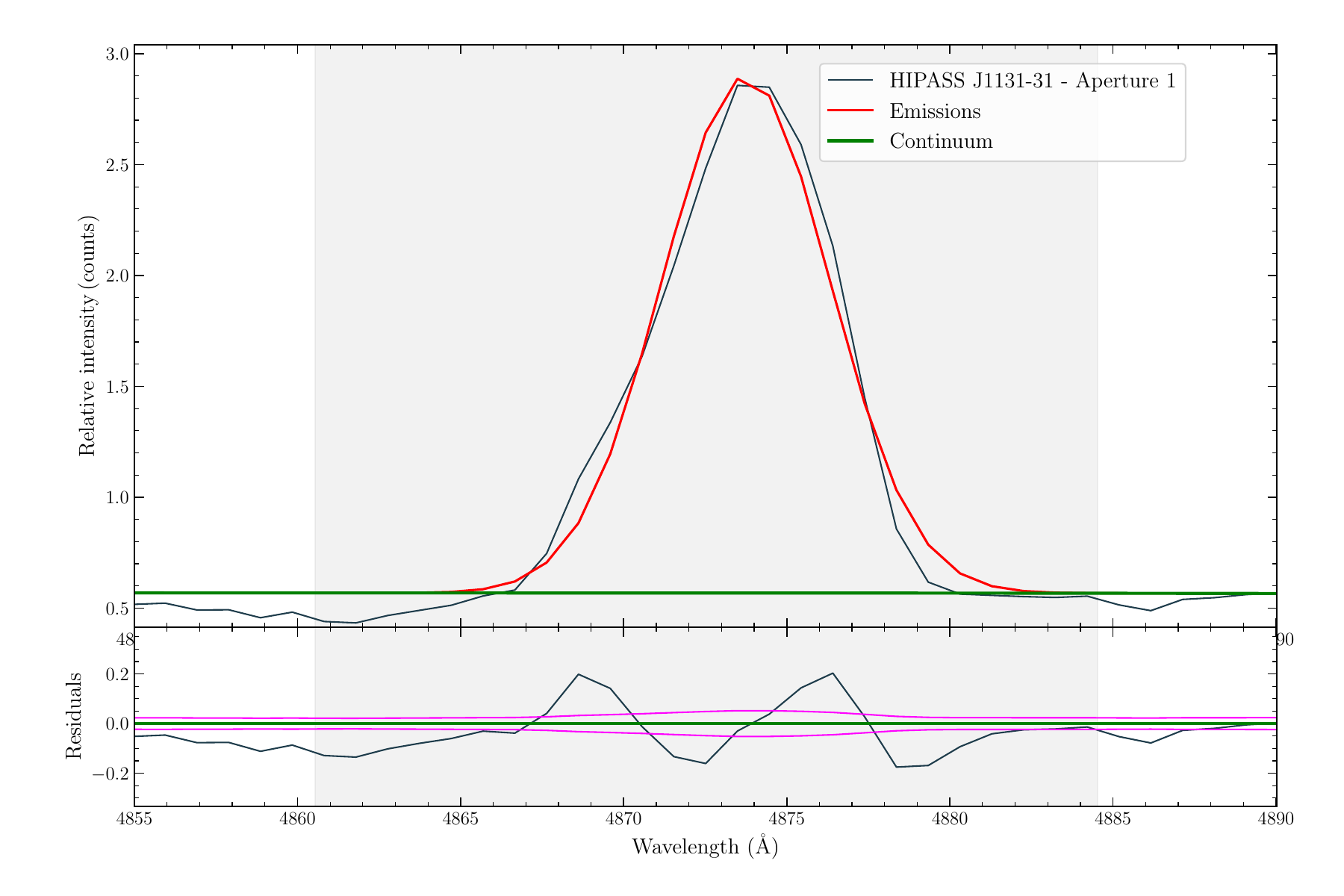}
        \includegraphics[clip=,angle=0,width=0.40\textwidth]{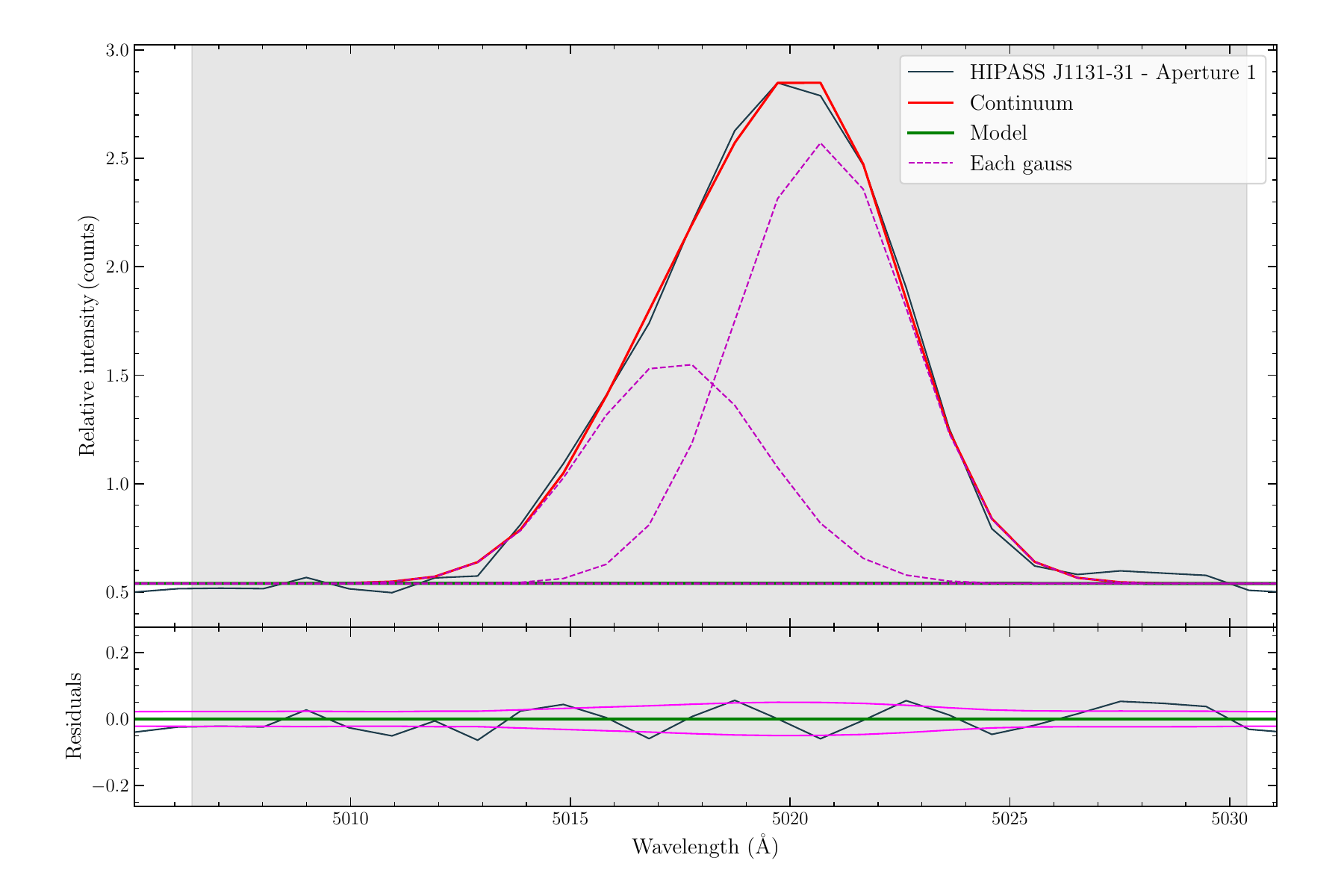}
        \includegraphics[clip=,angle=0,width=0.40\textwidth]{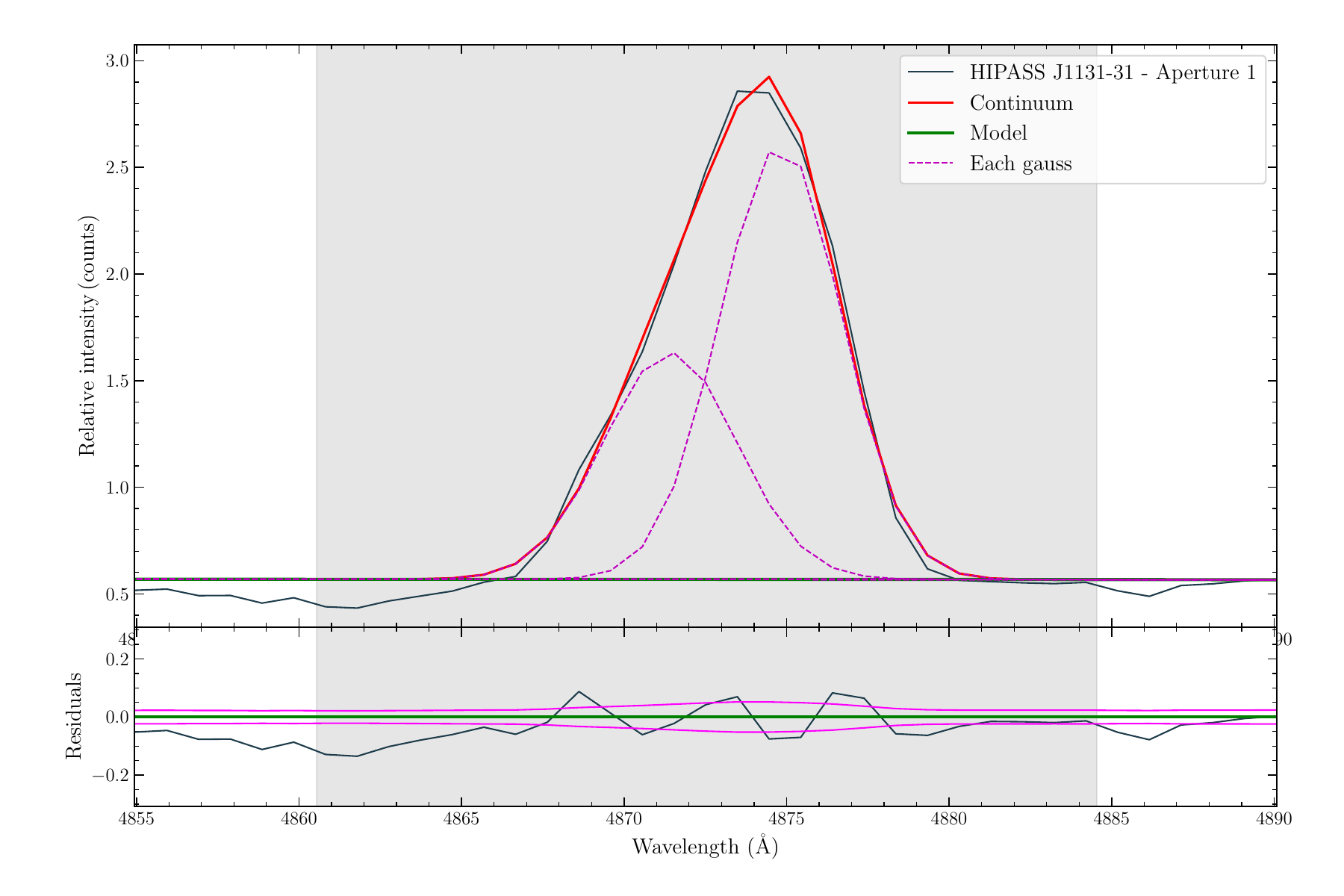}
    }
    \caption{Examples of one- and two-Gaussian fits of the emission lines H$\beta$ and [O{\sc iii}]$\lambda$5007
             in the spectrum of the E \HII\ region. The substantially reduced residual signal for the two-gauss
             fit is quantitatively described by the decreased several times the reduced $\chi$$^{2}$ statistic.
             See Sect.~\ref{ssec:HII-regions}.
        \label{fig:SALT_1d_E_rb}}
\end{figure}

\begin{figure}
    \centering{
        \includegraphics[clip=,angle=0,width=0.40\textwidth]{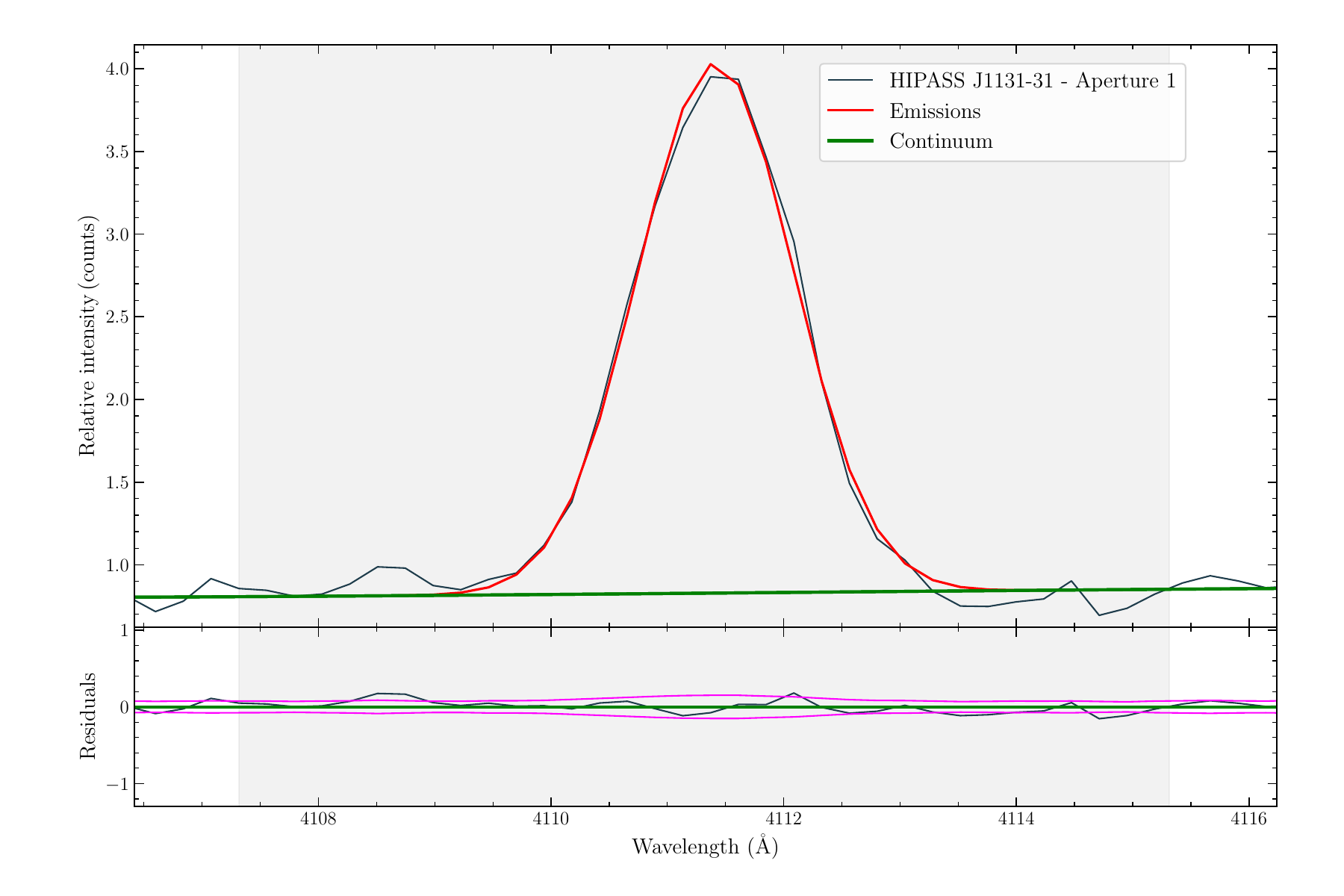}
        \includegraphics[clip=,angle=0,width=0.40\textwidth]{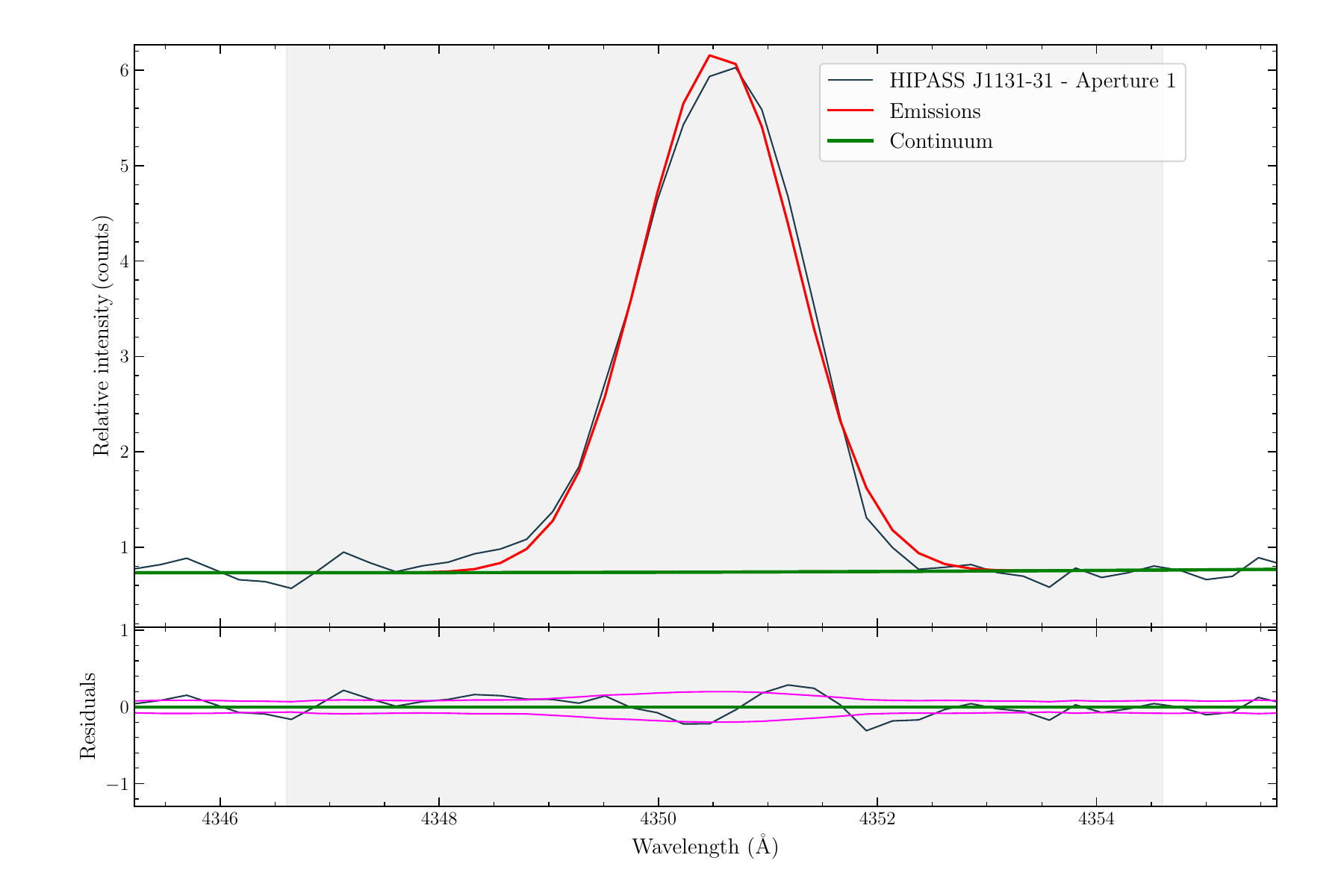}
        \includegraphics[clip=,angle=0,width=0.40\textwidth]{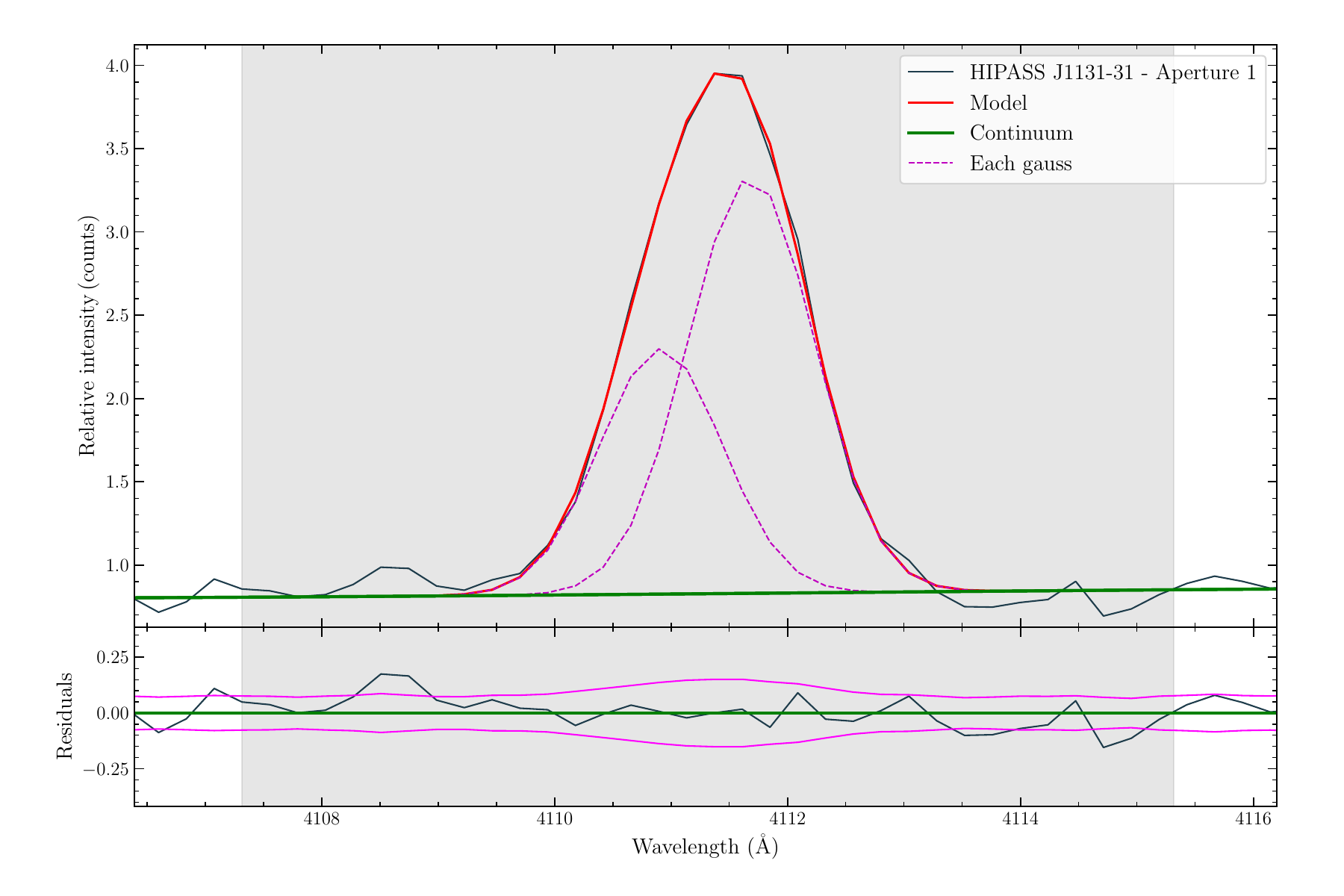}
        \includegraphics[clip=,angle=0,width=0.40\textwidth]{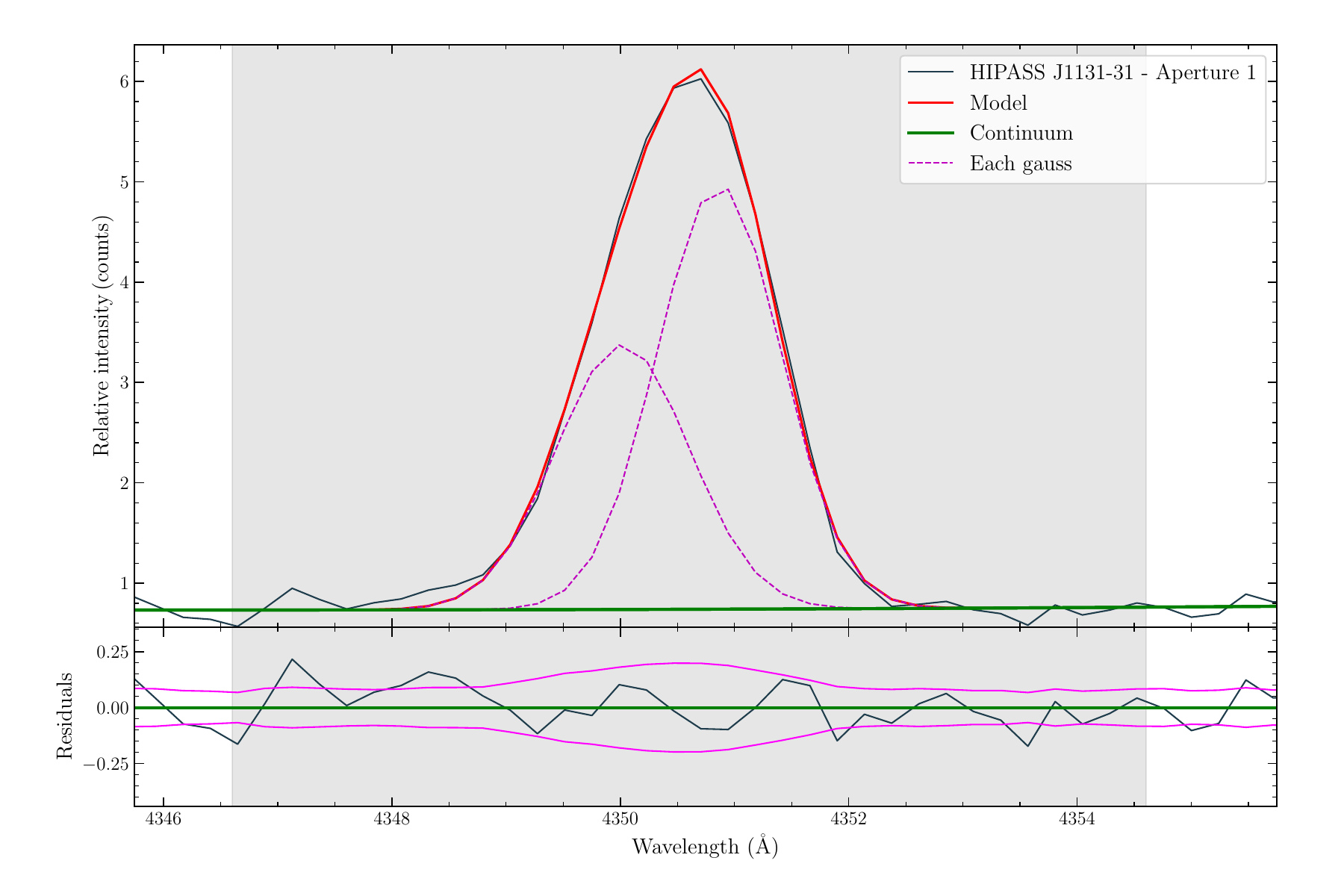}
    }
 \caption{Examples of one- and two-Gaussian fits of the emission lines H$\delta$ and H$\gamma$ observed with grating PG3000
	     in the spectrum of the E \HII\ region. The reduced $\chi^2$ statistic decreases by the factors of
	     approximately three  and two, respectively.
        \label{fig:SALT_1d_E_rb_PG3000}}
\end{figure}

\clearpage
\section{Line intensities and element abundances in the E and W \HII\ regions of the Peekaboo dwarf}
\label{sec:lines}

In Table~\ref{t:Intens} we present the measured relative line fluxes F($\lambda$) and corrected for extinction and underlying
Balmer absorptions relative line intensities I($\lambda$) for the three subsystems described in Sect.~\ref{ssec:SALT_data}
and \ref{ssec:HII-regions}. All details of line measurements and the subsequent determination of physical conditions
and element abundances are described in \citet{Sgr}. The relevant collision strengths and transition
probabilities were used  from  \citet{Stasinska05} and \citet{Izotov06}.
The results of the abundance calculations, obtained using these programmes, were compared with those derived from
the PyNeb package \citep{PyNeb}. The comparison demonstrates consistency within the stated uncertainties.

In the bottom of Table~\ref{t:Intens} we also present the derived extinction coefficient C(H$\beta$), equivalent widths
of Balmer absorptions EW(abs) in the underlying continuum and the equivalent widths of emission line H$\beta$ EW(H$\beta$).
These corrected line intensities I($\lambda$) are used for calculations of physical conditions and element abundances
according to recipes from \citet{Izotov06}. The standard two-zone model is assumed with the internal hot zone, where oxygen
is twice-ionised, and the outer colder zone, where oxygen is single-ionised.  In Table~\ref{t:Chem}, for the same three
subsystems, we present T$_{\rm e}$ for those two zones, and the estimate of electron density via the ratio of line fluxes in
[S{\sc ii}] doublet. Since for all three subsystems, this ratio exceeds the theoretical limit of 1.43, corresponding
to the limit of the low density (1~cm$^{-3}$), we adopt for all cases the conditional value of n$_{\rm e}$ = 10~cm$^{-3}$,
typical of majority of extragalactic \HII\ regions.
The uncertainties of T$_{\rm e}$ are calculated according to Equation 3 from \citet{Sgr} via the
standard error transfer from the known uncertainties of the related fluxes of the Oxygen lines. The temperatures
T$_{\rm e}$(O[{\sc ii}]) and T$_{\rm e}$(S[{\sc iii}]) are calculated using the approximation from \citet{Izotov06}.

The abundances of oxygen and other elements are derived with the direct method when the faint
auroral line [O\ {\sc iii}]$\lambda$4363 is available.
When this line is below the noise level (as in the subsystems of E(blue) and Region~W),
we  obtain the estimate of O/H with the modified semi-empirical method (mse) \citep[suggested in \citet{IT07} and
modified with the account for effect of parameter O$_{\mathrm {32}}$, in][]{BTA1}.
For the independent estimate of 12+log(O/H), we apply also the strong-line method of \citet{Izotov19DR14}, which appears
to demonstrate the best accuracy among all known empirical estimators. In particular, the independent check of its
accuracy on the extended sample of the low-metallicity \HII\ regions in \citet{BTA1}, resulted in its internal rms scatter
of only 0.04~dex. As one can see, the estimates of O/H with the three different methods for these \HII\ regions give
very close results.

The obtained abundance ratios of N/O, S/O, Ne/O, Ar/O in all three subsystems are consistent to each other.
  We can calculate their weighted averages as the characteristic values of the whole Peekaboo galaxy. The results are
  as follows: log(N/O)= --1.45$\pm$0.03, log(Ne/O)= --0.87$\pm$0.05; log(S/O) = -1.72$\pm$0.07, log(Ar/O) =
  --2.33$\pm$0.09. These averages are well consistent with the abundances ratios for the lowest metallicities, compiled
  by \citet[][in their Figure 10, left panel]{Izotov06}.

\begin{table*}[bhpt]
    \centering{
        \caption{Line intensities in the three subsystems of Peekaboo.}
        \label{t:Intens}
        \begin{tabular}{lcccccc} \hline
            \rule{0pt}{10pt}
            & \MC{2}{c}{subsystem E(red)} & \MC{2}{c}{subsystem E(blue)} & \MC{2}{c}{Region W}    \\ \hline
            \rule{0pt}{10pt}
            $\lambda_{0}$(\AA) Ion    & F($\lambda$)/F(H$\beta$) & I($\lambda$)/I(H$\beta$) & F($\lambda$)/F(H$\beta$) & I($\lambda$)/I(H$\beta$) & F($\lambda$)/F(H$\beta$) & I($\lambda$)/I(H$\beta$)\\ \hline\\[-0.30cm]
3727\ [O\ {\sc ii}]\      & 0.428$\pm$0.023 & 0.479$\pm$0.026 & 0.394$\pm$0.039 & 0.449$\pm$0.045 & 0.821$\pm$0.034 & 0.940$\pm$0.041 \\ 
3868\ [Ne\ {\sc iii}]\    & 0.088$\pm$0.009 & 0.098$\pm$0.009 & 0.081$\pm$0.017 & 0.090$\pm$0.019 & 0.062$\pm$0.013 & 0.070$\pm$0.015 \\ 
4101\ H$\delta$\          & 0.252$\pm$0.010 & 0.271$\pm$0.011 & 0.253$\pm$0.017 & 0.275$\pm$0.041 & 0.284$\pm$0.012 & 0.309$\pm$0.027 \\ 
4340\ H$\gamma$\          & 0.427$\pm$0.014 & 0.448$\pm$0.016 & 0.458$\pm$0.025 & 0.484$\pm$0.039 & 0.469$\pm$0.017 & 0.497$\pm$0.026 \\
4363\ [O\ {\sc iii}]\     & 0.032$\pm$0.005 & 0.033$\pm$0.005 & ---             & ---             & ---             & ---             \\ 
4861\ H$\beta$\           & 1.000$\pm$0.027 & 1.000$\pm$0.032 & 1.000$\pm$0.047 & 1.000$\pm$0.054 & 1.000$\pm$0.027 & 1.000$\pm$0.032 \\ 
4959\ [O\ {\sc iii}]\     & 0.330$\pm$0.022 & 0.327$\pm$0.022 & 0.370$\pm$0.037 & 0.366$\pm$0.037 & 0.231$\pm$0.020 & 0.227$\pm$0.019 \\ 
5007\ [O\ {\sc iii}]\     & 1.043$\pm$0.043 & 1.030$\pm$0.043 & 0.971$\pm$0.060 & 0.957$\pm$0.060 & 0.632$\pm$0.031 & 0.616$\pm$0.031 \\ 
5876\ He\ {\sc i}\        & 0.060$\pm$0.009 & 0.056$\pm$0.009 & 0.013$\pm$0.011 & 0.012$\pm$0.010 & 0.032$\pm$0.008 & 0.028$\pm$0.007 \\ 
6312\ [S\ {\sc iii}]\     & 0.003$\pm$0.002 & 0.003$\pm$0.002 & ---             & ---             & 0.008$\pm$0.003 & 0.007$\pm$0.003 \\ 
6563\ H$\alpha$\          & 3.096$\pm$0.059 & 2.740$\pm$0.058 & 3.126$\pm$0.106 & 2.714$\pm$0.101 & 3.139$\pm$0.104 & 2.714$\pm$0.098 \\ 
6584\ [N\ {\sc ii}]\      & 0.023$\pm$0.002 & 0.020$\pm$0.002 & 0.024$\pm$0.006 & 0.021$\pm$0.005 & 0.042$\pm$0.003 & 0.036$\pm$0.003 \\ 
6678\ He\ {\sc i}\        & 0.026$\pm$0.003 & 0.022$\pm$0.003 & 0.029$\pm$0.006 & 0.025$\pm$0.005 & 0.019$\pm$0.003 & 0.016$\pm$0.003 \\ 
6717\ [S\ {\sc ii}]\      & 0.046$\pm$0.004 & 0.041$\pm$0.003 & 0.037$\pm$0.006 & 0.032$\pm$0.005 & 0.070$\pm$0.005 & 0.060$\pm$0.004 \\ 
6731\ [S\ {\sc ii}]\      & 0.021$\pm$0.003 & 0.018$\pm$0.003 & 0.026$\pm$0.006 & 0.022$\pm$0.005 & 0.038$\pm$0.003 & 0.033$\pm$0.003 \\
7065\ He\ {\sc i}\        & 0.025$\pm$0.003 & 0.022$\pm$0.002 & ---             & ---             & 0.019$\pm$0.003 & 0.016$\pm$0.003 \\ 
7136\ [Ar\ {\sc iii}]\    & 0.020$\pm$0.003 & 0.017$\pm$0.003 & 0.009$\pm$0.004 & 0.007$\pm$0.003 & 0.011$\pm$0.003 & 0.009$\pm$0.003 \\
  & & \\
C(H$\beta$)\ dex          & \MC {2}{c}{0.16$\pm$0.03}         & \MC {2}{c}{0.18$\pm$0.04}         & \MC {2}{c}{0.19$\pm$0.04}         \\  
EW(abs)\ \AA\             & \MC {2}{c}{0.00$\pm$0.29}         & \MC {2}{c}{0.00$\pm$0.26}         & \MC {2}{c}{0.00$\pm$0.33}         \\   
EW(H$\beta$)\ \AA\        & \MC {2}{c}{  18$\pm$ 2}           & \MC {2}{c}{  12$\pm$ 3}           & \MC {2}{c}{  19$\pm$3}            \\  
            \hline
        \end{tabular}
    }
\end{table*}

\begin{table*}
    \centering{
        \caption{Abundances in three components of Peekaboo.}
        \label{t:Chem}
        \begin{tabular}{lccc} \hline
            \rule{0pt}{10pt}
            Value                              & subsystem E(red)(4363)& subsystem E(blue)$\dagger$  & Region W$\dagger$    \\
                                               & PG900+PG3000+PG1800   & PG900+PG3000+PG1800   & PG900+PG3000+PG1800  \\ \hline\\[-0.30cm]
            $T_{\rm e}$(OIII)(K)\              & 19,682$\pm$1990~~     &  22,032$\pm$2316~~    & 22,488$\pm$2285~~    \\
            $T_{\rm e}$(OII)(K)\               & 15,582$\pm$208 ~~     &  16,209$\pm$508 ~~    & 16,305$\pm$463 ~~    \\
            $T_{\rm e}$(SIII)(K)\              & 18,542$\pm$1327~~     &  19,877$\pm$1087~~    & 20,082$\pm$985 ~~    \\
            $N_{\rm e}$(SII)(cm$^{-3}$)\       &  10$\pm$10 ~~         &   10$\pm$10 ~~        &  10$\pm$10 ~~        \\
            & \\                                                       
            O$^{+}$/H$^{+}$($\times$10$^5$)\   & 0.387$\pm$0.026~~     &  0.323$\pm$0.044~~    & 0.664$\pm$0.062~~    \\
            O$^{++}$/H$^{+}$($\times$10$^5$)\  & 0.601$\pm$0.128~~     &  0.468$\pm$0.098~~    & 0.289$\pm$0.057~~    \\
            O/H($\times$10$^5$)\               & 0.988$\pm$0.131~~     &  0.790$\pm$0.107~~    & 0.953$\pm$0.084~~    \\
            12+log(O/H) (dir)\                 & ~6.99$\pm$0.06~~      &  ...                  &  ...                 \\
            12+log(O/H)(mse,c)\                & ~6.96$\pm$0.11~~      &  ~6.94$\pm$0.11~~     & ~7.02$\pm$0.10~~     \\
                12+log(O/H)(s)\                & ~7.01$\pm$0.04~~      &  ~6.98$\pm$0.05~~     & ~7.02$\pm$0.05~~      \\
            & \\                                                       
            N$^{+}$/H$^{+}$($\times$10$^7$)\   & 1.429$\pm$0.121~~     &  1.343$\pm$0.283~~    &  2.304$\pm$0.202~~  \\
            ICF(N)\                            & 2.573                 &  2.471                &  1.368              \\
            N/H($\times$10$^5$)\               & 0.04$\pm$0.00~~       &  0.03$\pm$0.01~~      &  0.03$\pm$0.00~~    \\
            12+log(N/H)\                       & 5.57$\pm$0.04~~       &  5.52$\pm$0.09~~      &  5.50$\pm$0.04~~    \\
            log(N/O)\                          & -1.43$\pm$0.07~~      &  -1.38$\pm$0.11~~     &  -1.48$\pm$0.05~~   \\
            & \\                                                                               
            Ne$^{++}$/H$^{+}$($\times$10$^5$)\ & 0.128$\pm$0.033~~     &  0.092$\pm$0.028~~    &  0.068$\pm$0.020~~  \\
            ICF(Ne)\                           & 1.167                 &  1.174                &  1.321              \\
            Ne/H($\times$10$^5$)\              & 0.15$\pm$0.04~~       &  0.11$\pm$0.03~~      &  0.09$\pm$0.03~~    \\
            12+log(Ne/H)\                      & 6.17$\pm$0.11~~       &  6.03$\pm$0.13~~      &  5.96$\pm$0.13~~    \\
            log(Ne/O)\                         & -0.82$\pm$0.12~~      &  -0.86$\pm$0.15~~     &  -1.02$\pm$0.14~~   \\
            & \\                                                                               
            S$^{+}$/H$^{+}$($\times$10$^7$)\   & 0.529$\pm$0.040~~     &  ---                  &  0.772$\pm$0.057~~  \\
            S$^{++}$/H$^{+}$($\times$10$^7$)\  & 0.846$\pm$0.553~~     &  ---                  &  1.620$\pm$0.679~~  \\
            ICF(S)\                            & 0.970                 &  ---                  &  0.826              \\
            S/H($\times$10$^7$)\               & 1.33$\pm$0.54~~       &  ---                  &  1.98$\pm$0.56~~    \\
            12+log(S/H)\                       & 5.12$\pm$0.18~~       &  ---                  &  5.30$\pm$0.12~~    \\
            log(S/O)\                          & -1.87$\pm$0.18~~      &  ---                  &  -1.68$\pm$0.13~~   \\
            & \\                                                                               
            Ar$^{++}$/H$^{+}$($\times$10$^7$)\ & 0.489$\pm$0.092~~     &  0.196$\pm$0.080~~    &  0.234$\pm$0.069~~  \\
            ICF(Ar)\                           & 1.075                 &  1.074                &  1.103              \\
            Ar/H($\times$10$^7$)\              & 0.53$\pm$0.10~~       &  0.21$\pm$0.09~~      &  0.26$\pm$0.08~~    \\
            12+log(Ar/H)\                      & 4.72$\pm$0.08~~       &  4.32$\pm$0.18~~      &  4.41$\pm$0.13~~    \\
            log(Ar/O)\                         & -2.27$\pm$0.10~~      &  -2.58$\pm$0.19~~     &  -2.57$\pm$0.13~~   \\
            \hline
\\[-0.35cm]
\multicolumn{4}{l}{$\dagger$ Abundances of O, N, Ne, and Ar for subsystems E(blue) and Region W, and of S for Region W are calculated with T$_{\rm e}$,} \\
\multicolumn{4}{l}{adopted in the modified semi-empirical method (mse) \citep{BTA1}. Relative values X/O are direct results of calculations.} \\
\multicolumn{4}{l}{12+log(O/H)(mse) is, however, corrected upward by 0.04~dex for consistency with O/H(T$_{\rm e}$) zero point  \citep[see][]{BTA1}.} \\
        \end{tabular}
    }
\end{table*}

\clearpage
\section{HST data: Identification of the XMP massive hot stars and supergiants}
\label{sec:HST_stars}

\begin{figure*}[bhpt]
    \centering{
	\includegraphics[clip=,angle=-0,width=0.70\textwidth]{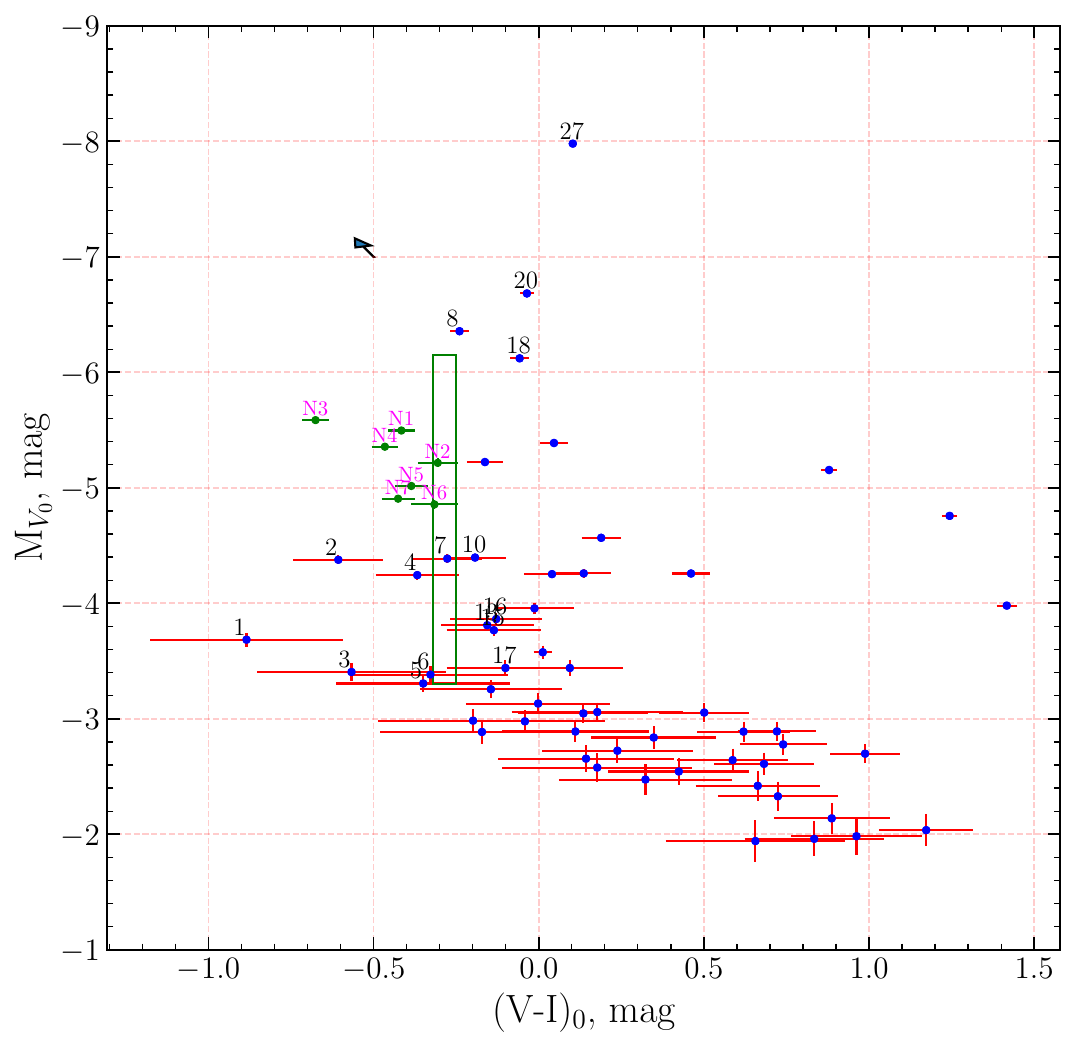}
    }
    \caption{Hubble Space Telescope-based diagram of 'M$_{\rm V}$ versus ($V-I$)', corrected for MW extinction, for all stars within
    the Peekaboo dwarf body.  Several the hottest stars ($(V-I)_{\mathrm 0} <$ --0.25 mag) and four supergiant stars
    (M$_{\rm V,0} <$ --6.0~mag)  are marked by their numbers in Tab.~\ref{tab:all-stars}.
    The additional extinction of $\delta$A$_{\rm V} =$0.16~mag and reddening of $\delta$($V-I$) = 0.057~mag
    in the E and W  \HII-regions (estimated above in this section) is not taken into account in the plot. Its
    effect is shown by an arrow in the upper-left section of the plot.   Stars No.~1, 2 and 3 show very
    blue $(V-I)_{\mathrm 0}$ of --0.88 and $\sim$--0.60. Despite rather large errors $\sigma$($V-I$),
    they are much bluer of the border of the main sequence at (V--I) of the hottest OV-stars of --0.32~mag \citep{Lorenzo25}.
    These 'blue outliers' can be very hot evolved stars, with the surface temperatures up to 150--200~kK as suggested
    by \citet{Lorenzo22} for several such stars in the galaxy Sextans~A. We plot also positions of 5 'nebulosities' (N1--N5)
    within the borders of W and E \HII\ regions, which are  marked by circles in Fig.~\ref{fig:HST_color} (top panel).
        \label{fig:cmd}}
\end{figure*}

We plot in Fig.~\ref{fig:cmd} positions of all HST stars within the Peekaboo body in the
diagram M$_{\rm V}$ versus colour $(V-I)_{\mathrm 0}$. We mark by their numbers from Table~\ref{tab:all-stars} all
'blue' hot stars and the four
supergiants with M$_{\rm V} <$ --6.0~mag. Here, for both M$_{\rm V}$ and $(V-I)_{\mathrm 0}$, only extinction in the Galaxy
is accounted for, corresponding to E(B-V) = 0.062 \citep{SF2011}, and the related A$_{\rm V}$ = 0.186~mag and
E($V-I$) = 0.066~mag.
As one can see in the bottom of Table~\ref{t:Intens}, C(H$\beta$) for all three systems is the same within rather small
uncertainties. We adopt its average value of 0.17. Then, for stars residing within the borders of these \HII\ regions,
one can expect the elevated extinction and reddening, in addition to the adopted Milky Way values,
$\delta$A$_{\rm V}$ = 0.16~mag and  $\delta$E($V-I$)=0.057~mag.  In fact, we do not know how this extinction is
distributed within the E and W regions. If it is localised within the 'blue nebulosities', then this 'reddening' correction
is probably applicable only to stars residing close to these 'blue nebulosities'.

Therefore, looking at the marked positions of blue and bluish stars in the Peekaboo dwarf image in
Fig.~\ref{fig:HST_color} (top), one can understand where these stars shift in Fig.~\ref{fig:cmd} with the account of
the additional extinction (its value is shown by the "arrow" in the plot). Curiously, but no stellar objects are
identified within the W region. At the same time, two very blue and rather luminous 'nebulosities', N4 and N5 can
be the main contributors to the visible nebular emission of the W \HII\ region.
As for the E \HII\ region, there are several stars, for which the additional extinction can take place. These are stars
No. 1, 2, 4, 5, 6, 7, 8. Stars No. 10, 15, 18 and 20 reside in the 'narrow' layer between the E and W \HII\ regions and
also can be affected by the additional extinction. So, for some of the discussed blue stars, their colour index $(V-I)$
can be bluer than that shown in Fig.~\ref{fig:cmd}. This implies that some of visible hot stars within the E region
can shift after the colour correction to the area of 'blue outliers', where the very hot helium burning stars similar
to WO stars reside \citep{Lorenzo22}.

\begin{table*}
    \centering{
    \caption{Magnitudes and colours of the HST stars within the Peekaboo dwarf body, partly adopted from the EDD table.}
\label{tab:all-stars}
    \begin{tabular}{llllccclcc} \hline\\[-0.30cm]
 id  & $V-I$   & $V$    & $V_{\rm err}$&$I_{\rm err}$ &M$_{\rm V,0}$&$(V-I)_{\rm 0}$ &$(V-I)_{\rm er}$ & x       & y  \\
     & mag     & mag    & mag         & mag         & mag         & mag            & mag    & px      & px \\
        \hline\\[-0.30cm]
01   & -0.819  & 25.701 & 0.060       & 0.206       & -3.685      & -0.885 &  0.291   & 1945.99 & 2406.69   \\
02   & -0.541  & 25.009 & 0.038       & 0.096       & -4.377      & -0.607 &  0.136   & 1938.69 & 2411.02   \\
03   & -0.501  & 25.981 & 0.074       & 0.202       & -3.405      & -0.567 &  0.286   & 1908.78 & 2438.43   \\
04   & -0.302  & 25.142 & 0.039       & 0.089       & -4.244      & -0.368 &  0.126   & 1926.85 & 2396.91   \\
05   & -0.284  & 26.080 & 0.078       & 0.186       & -3.306      & -0.350 &  0.263   & 1922.01 & 2382.68   \\
06   & -0.262  & 26.003 & 0.072       & 0.166       & -3.383      & -0.328 &  0.235   & 1919.79 & 2394.47   \\
07   & -0.211  & 24.999 & 0.036       & 0.075       & -4.387      & -0.277 &  0.106   & 1954.70 & 2435.43   \\
08   & -0.174  & 23.030 & 0.012       & 0.020       & -6.356      & -0.240 &  0.028   & 1932.66 & 2406.53   \\
09   & -0.133  & 26.402 & 0.098       & 0.204       & -2.984      & -0.199 &  0.288   & 1979.08 & 2378.20   \\
10   & -0.127  & 24.990 & 0.034       & 0.067       & -4.396      & -0.193 &  0.095   & 1900.32 & 2388.97   \\
11   & -0.106  & 26.500 & 0.101       & 0.218       & -2.886      & -0.172 &  0.308   & 1936.33 & 2336.98   \\
12   & -0.097  & 24.163 & 0.021       & 0.039       & -5.223      & -0.163 &  0.055   & 1907.77 & 2413.47   \\
13   & -0.090  & 25.576 & 0.052       & 0.099       & -3.810      & -0.156 &  0.140   & 1906.00 & 2448.44   \\
14   & -0.079  & 26.130 & 0.077       & 0.152       & -3.256      & -0.145 &  0.215   & 1903.38 & 2446.73   \\
15   & -0.070  & 25.619 & 0.051       & 0.101       & -3.767      & -0.136 &  0.143   & 1903.17 & 2373.19   \\
16   & -0.063  & 25.522 & 0.051       & 0.098       & -3.864      & -0.129 &  0.139   & 1900.36 & 2407.79   \\
17   & -0.035  & 25.946 & 0.066       & 0.126       & -3.440      & -0.101 &  0.178   & 1897.10 & 2412.65   \\
18   &  0.008  & 23.265 & 0.012       & 0.020       & -6.121      & -0.058 &  0.028   & 1917.10 & 2370.44   \\
19   &  0.024  & 26.408 & 0.094       & 0.172       & -2.978      & -0.042 &  0.243   & 1899.36 & 2425.37   \\
20   &  0.030  & 22.703 & 0.009       & 0.014       & -6.683      & -0.036 &  0.020   & 1925.07 & 2369.57   \\
21   &  0.053  & 25.430 & 0.048       & 0.084       & -3.956      & -0.013 &  0.119   & 1951.46 & 2410.61   \\
22   &  0.064  & 26.254 & 0.092       & 0.155       & -3.132      & -0.002 &  0.219   & 1941.83 & 2406.24   \\
23   &  0.078  & 25.810 & 0.057       & 0.020       & -3.576      &  0.012 &  0.028   & 1898.75 & 2458.86   \\
24   &  0.106  & 25.133 & 0.037       & 0.060       & -4.253      &  0.040 &  0.085   & 1921.18 & 2411.16   \\
25   &  0.112  & 23.998 & 0.018       & 0.029       & -5.388      &  0.046 &  0.041   & 1909.35 & 2342.51   \\
26   &  0.160  & 25.946 & 0.071       & 0.113       & -3.440      &  0.094 &  0.160   & 1953.89 & 2439.19   \\
27   &  0.169  & 21.406 & 0.004       & 0.006       & -7.980      &  0.103 &  0.008   & 1942.40 & 2466.17   \\
28   &  0.177  & 26.496 & 0.095       & 0.157       & -2.890      &  0.111 &  0.222   & 2011.78 & 2403.17   \\
29   &  0.201  & 26.339 & 0.086       & 0.139       & -3.047      &  0.135 &  0.197   & 1940.86 & 2340.86   \\
30   &  0.202  & 25.127 & 0.039       & 0.059       & -4.259      &  0.136 &  0.083   & 1911.97 & 2413.27   \\
31   &  0.209  & 26.731 & 0.117       & 0.188       & -2.655      &  0.143 &  0.266   & 1921.02 & 2344.43   \\
32   &  0.243  & 26.328 & 0.083       & 0.183       & -3.058      &  0.177 &  0.259   & 1875.11 & 2401.46   \\
33   &  0.243  & 26.808 & 0.124       & 0.204       & -2.578      &  0.177 &  0.288   & 1896.78 & 2302.76   \\
34   &  0.255  & 24.819 & 0.028       & 0.042       & -4.567      &  0.189 &  0.059   & 1893.44 & 2308.65   \\
35   &  0.304  & 26.662 & 0.108       & 0.161       & -2.724      &  0.238 &  0.228   & 1904.78 & 2340.19   \\
36   &  0.389  & 26.912 & 0.131       & 0.185       & -2.474      &  0.323 &  0.262   & 1941.23 & 2345.01   \\
37   &  0.414  & 26.548 & 0.098       & 0.134       & -2.838      &  0.348 &  0.189   & 1923.55 & 2384.49   \\
38   &  0.490  & 26.842 & 0.115       & 0.151       & -2.544      &  0.424 &  0.214   & 1878.62 & 2421.48   \\
39   &  0.527  & 25.128 & 0.034       & 0.041       & -4.258      &  0.461 &  0.058   & 1913.15 & 2427.36   \\
40   &  0.567  & 26.332 & 0.081       & 0.097       & -3.054      &  0.501 &  0.137   & 1905.14 & 2404.46   \\
41   &  0.653  & 26.744 & 0.100       & 0.118       & -2.642      &  0.587 &  0.167   & 1900.86 & 2295.04   \\
42   &  0.686  & 26.497 & 0.086       & 0.099       & -2.889      &  0.620 &  0.140   & 1933.49 & 2343.37   \\
43   &  0.722  & 27.444 & 0.178       & 0.192       & -1.942      &  0.656 &  0.272   & 1882.12 & 2400.30   \\
44   &  0.729  & 26.966 & 0.129       & 0.133       & -2.420      &  0.663 &  0.188   & 1925.61 & 2392.75   \\
45   &  0.748  & 26.776 & 0.098       & 0.106       & -2.610      &  0.682 &  0.150   & 1982.51 & 2436.69   \\
46   &  0.787  & 26.495 & 0.082       & 0.084       & -2.891      &  0.721 &  0.119   & 1873.15 & 2391.30   \\
47   &  0.790  & 27.057 & 0.127       & 0.128       & -2.329      &  0.724 &  0.181   & 1868.12 & 2413.24   \\
48   &  0.806  & 26.608 & 0.088       & 0.093       & -2.778      &  0.740 &  0.132   & 1977.93 & 2313.70   \\
49   &  0.900  & 27.425 & 0.153       & 0.149       & -1.961      &  0.834 &  0.211   & 1854.12 & 2307.70   \\
50   &  0.945  & 24.232 & 0.017       & 0.016       & -5.154      &  0.879 &  0.023   & 1936.36 & 2421.46   \\
51   &  0.953  & 27.248 & 0.134       & 0.124       & -2.138      &  0.887 &  0.175   & 1869.25 & 2406.83   \\
52   &  1.028  & 27.403 & 0.158       & 0.139       & -1.983      &  0.962 &  0.197   & 1988.41 & 2337.56   \\
53   &  1.054  & 26.688 & 0.082       & 0.074       & -2.698      &  0.988 &  0.105   & 1853.88 & 2431.71   \\
54   &  1.239  & 27.350 & 0.140       & 0.101       & -2.036      &  1.173 &  0.143   & 1865.83 & 2343.97   \\
55   &  1.310  & 24.629 & 0.020       & 0.016       & -4.757      &  1.244 &  0.023   & 1931.53 & 2330.40   \\
56   &  1.483  & 25.406 & 0.033       & 0.022       & -3.980      &  1.417 &  0.031   & 1920.05 & 2350.67   \\
        \hline
     \end{tabular}
   }
 \end{table*}

\begin{table*}
  \centering{
    \caption{Magnitudes and colours of the HST brightest \HII\ regions within the Peekaboo body.}
    \label{tab:all-HII}
    \begin{tabular}{cccccccccc} \hline\\[-0.30cm]
 id  & $V-I$    & $V$   &$V_{\rm err}$&$I_{\rm err}$ &M$_{\rm V,0}$&$(V-I)_{\rm 0}$ &$(V-I)_{\rm er}$ & RA     & DEC         \\
     & mag         & mag   & mag      & mag          & mag        & mag            & mag       & degrees       & degrees     \\
\hline\\[-0.30cm]
N\,1    &-0.35     & 23.89 &  0.02    & 0.03         & -5.496     & -0.416         & 0.04      &  172.8952917  & -31.6739733 \\
N\,2    &-0.24     & 24.17 &  0.04    & 0.05         & -5.216     & -0.306         & 0.06      &  172.8947500  & -31.6739306 \\
N\,3    &-0.61     & 23.80 &  0.02    & 0.03         & -5.586     & -0.676         & 0.04      &  172.8952458  & -31.6744069 \\
N\,4    &-0.40     & 24.03 &  0.02    & 0.04         & -5.356     & -0.466         & 0.04      &  172.8940750  & -31.6740158 \\
N\,5    &-0.32     & 24.37 &  0.03    & 0.04         & -5.016     & -0.386         & 0.05      &  172.8936750  & -31.6741089 \\
N\,6    &-0.25     & 24.53 &  0.04    & 0.06         & -4.856     & -0.316         & 0.07      &  172.8950989  & -31.6745340 \\
N\,7    &-0.36     & 24.48 &  0.03    & 0.04         & -4.906     & -0.426         & 0.05      &  172.8945400  & -31.6746452 \\
\hline                                                                         
     \end{tabular}
   }
 \end{table*}

\begin{table*}
  \centering{
  \caption{General properties of known XMP galaxies within the LV.}
    \label{tab:XMP-LV}
    \begin{tabular}{llllcrllc} \hline\\[-0.30cm]
Name          & Alter.name & 12+log(O/H) & Method       &V$_{\rm hel}$ & $D$        &M$_{\rm B,0}$& Coord. J2000    & Ref.     \\
              &            & dex         &              & \kms         & Mpc        & mag         &                 & of (O/H)  \\
\hline\\[-0.30cm]
HIPASSJ1131-31 & Peekaboo &  6.99$\pm$0.06& T$_{\rm e}$ & 716          & 6.8        & --11.25      & J113134.6-314026&   1        \\
AGC227973      &          &  7.07$\pm$0.04& s.l.        & 675          & 9.1        & --10.42      & J125039.9+052052&   2        \\
SDSSJ0926+3343 & AGC194054&  7.12$\pm$0.02& T$_{\rm e}$ & 532          & 10.8       & --12.92      & J092610.1+334312&   3        \\
SDSSJ1444+4242 &          &  7.14$\pm$0.04& s.l.        & 634          & 10.9       & --10.54      & J144449.8+424254&   4        \\
PGC000083      & AGC103567&  7.15$\pm$0.03& s.l.        & 542          & 9.1        & --12.89      & J000106.5+322241&   4        \\
SDSSJ1056+3608 & AGC205685&  7.16$\pm$0.07& T$_{\rm e}$ & 572          & 9.2        & --11.60      & J105640.4+360828&   5        \\
HIJ1021+1805   & Leo~P    &  7.17$\pm$0.04& T$_{\rm e}$ & 264          & 1.6        & --09.11      & J102144.8+180520&   6        \\
\hline                                                                         
\multicolumn{9}{l}{1. This paper; 2. \citet{SALT2}; 3. \citet{J0926}; } \\
\multicolumn{9}{l}{4. \citet{BTA1}; 5. \citet{ITG12}; 6. \citet{Skillman13}} \\
     \end{tabular}
   }
 \end{table*}
\end{appendix}

\begin{thebibliography}{99}

\bibitem[\protect\citeauthoryear{Anand et al.}{2021}]{Anand21}
Anand G.S., Rizzi L., Tully R.B., et al., 2021, AJ, 162, 80 (15pp)

\bibitem[\protect\citeauthoryear{Anders \& Fritze-v.~Alvensleben }{2003}]{Anders03}
Anders P. \& Fritze-v.~Alvensleben U., 2003, A\&A, 401, 1063

\bibitem[\protect\citeauthoryear{Annibali et al.}{2019}]{Annibali2019}
        Annibali F., La Torre V., Tosi M., et al., 2019, MNRAS, 482, 3892

\bibitem[\protect\citeauthoryear{Buckley, Swart \& Meiring}{2006}]{Buck06}
Buckley, D.A.H., Swart, G.P., Meiring, J.G., 2006, SPIE, 6267

\bibitem[\protect\citeauthoryear{Burgh et al.}{2003}]{Burgh03}
Burgh, E.B., Nordsieck, K.H., Kobulnicky, H.A., Williams, T.B., O'Donoghue, D., Smith, M.P., Percival, J.W.,
2003, SPIE, 4841, 1463

\bibitem[\protect\citeauthoryear{Crawford et al.}{2010}]{Cra2010}
Crawford S.M. et al., 2010, in Silva D. R., Peck A. B., Soifer B. T.,
Proc. SPIE Conf. Ser. Vol. 7737, Observatory Operations:
Strategies, Processes, and Systems III. SPIE, Bellingham, p. 773725

\bibitem[\protect\citeauthoryear{Crowther \& Bibby}{2009}]{Crowther2009}
  Crowther P.A., Bibby J.L., 2009, A\&A, 499, 455

\bibitem[\protect\citeauthoryear{Eldridge, Stanway}{2022}]{Eldridge22}
  Eldridge J.J., Stanway E.E., 2022, ARAA, 60, 455

\bibitem[\protect\citeauthoryear{Garcia et al.}{2021}]{Garcia21}
  Garcia M., Evans C.J., Bestenlehner J.M. et al., 2021, Experimental Astronomy, 51:887-911

\bibitem[\protect\citeauthoryear{Gull et al.}{2022}]{Gull22}
  Gull M., Weisz D.R., Senchyna P., et al., 2022, ApJ, 941, 206

\bibitem[\protect\citeauthoryear{Izotov et al.}{2006}]{Izotov06}
Izotov Y.I., Stasi\'{n}ska G., Meynet G., et al., 2006, A\&A, 448, 955

\bibitem[\protect\citeauthoryear{Izotov \& Thuan}{2007}]{IT07}
Izotov Y.I., Thuan T.X., 2007,  \apj,  665, 1115


\bibitem[\protect\citeauthoryear{Izotov, Thuan, \& Guseva}{Izotov et al.}{2007}]{ITG07}
Izotov Y.I., Thuan T.X., Guseva N.G., 2007, \apj, 671, 1297

\bibitem[\protect\citeauthoryear{Izotov, Thuan, Guseva}{Izotov et al.}{2012}]{ITG12}
Izotov Y.I., Thuan T.X., Guseva N.G., 2012,  A\&A,  546, A122

\bibitem[\protect\citeauthoryear{Izotov et al.}{2019}]{Izotov19DR14}
Izotov Y.I., Guseva N.G., Frieke K.J., Henkel C., 2019, A\&A, 523, A40

\bibitem[\protect\citeauthoryear{Karachentsev et al.}{2023}]{Peekaboo}
Karachentsev I.D., Makarova L.N., Koribalski B.S., Anand G.S., Tully R.B., Kniazev A.Y.,
  2023, MNRAS, 518, 5893

\bibitem[\protect\citeauthoryear{Kniazev}{2025}]{Kniazev2025}
Kniazev A.Y., 2025, RAA, 25, id.045012, 16 pp.

\bibitem[\protect\citeauthoryear{Kniazev et al.}{2008}]{Sgr}
Kniazev A.Y., et al.\,  2008, MNRAS, 388, 1667

\bibitem[\protect\citeauthoryear{Kniazev}{2022}]{Kniazev2022}
Kniazev A.Y., 2022, Astrophysical Bulletin, 77, 334-346

\bibitem[\protect\citeauthoryear{Kobulnicky et al.}{2003}]{Kobul03}
Kobulnicky H.A., Nordsieck K.H., Burgh E.B., Smith M.P., Percival J.W., Williams T.B., O'Donoghue D.,
2003, SPIE, 4841, 1634

\bibitem[\protect\citeauthoryear{Koribalski et al.}{2018}]{Koribalski2018}
Koribalski B.S., Wang J., Kamphius P., et al.
  2018, MNRAS, 478, 1611

\bibitem[\protect\citeauthoryear{Li et al.}{2024}]{Legacy}
Li C., Zhang Y., Cui C., Wei S., Zhang J., Zhao Y., Wu X.-B., et al.,
2024, AJ, 168, 233

\bibitem[\protect\citeauthoryear{Lorenzo et al.}{2022}]{Lorenzo22}
   Lorenzo M., Garcia M., Najarro F., Herrero A., Cervi\~no M., Castro N.,
   2022, MNRAS, 516, 4164

\bibitem[\protect\citeauthoryear{Lorenzo et al.}{2025}]{Lorenzo25}
Lorenzo M., Garcia M., Castro N., Najarro F., Cervi\~{n}o M., Herrero A., Sim\'{o}n-D\'{i}az S.,
2025, MNRAS, 537, 1197

\bibitem[\protect\citeauthoryear{Lupton et al.}{2005}]{Lupton05}
 \mbox{Lupton~R.,~et~al.~2005}, https://www.sdss3.org/dr8/algorithms/ \\sdssUBVRITransform.php\#Lupton2005

\bibitem[\protect\citeauthoryear{Luridiana, Morisset, \& Shaw}{2015}]{PyNeb}
Luridiana V., Morisset C., Shaw R.~A., 2015, A\&A, 573, A42

\bibitem[\protect\citeauthoryear{McQuinn et al.}{2012}]{McQuinn12}
McQuinn K.B.W., Skillman E.D., Dalcanton J.J., Cannon J.M., Dolphin A.E., Holtzman J., Weisz D.R., Williams B.F.,
2012, ApJ, 759, id.77 (17pp)

\bibitem[\protect\citeauthoryear{O'Donoghue et al.}{2006}]{Dono06}
O'Donoghue D., et al.\, 2006, \mnras, 372, 1510


\bibitem[\protect\citeauthoryear{Pustilnik, Kniazev, \& Pramskij}{Pustilnik et al}{2005}]{DDO68}
  Pustilnik S., Kniazev A., Pramskij A., 2005, \aap,  443, 91

\bibitem[\protect\citeauthoryear{Pustilnik et al.}{2010}]{J0926}
Pustilnik S.A., Tepliakova A.L., Kniazev A.Y., Burenkov A.N., 2010, MNRAS, 401, 333


\bibitem[\protect\citeauthoryear{Pustilnik, Tepliakova, Makarov}{Pustilnik et al.}{2019}]{PTM19}
Pustilnik S.A., Tepliakova A.L., Makarov D.I., 2019, MNRAS, 482, 4329

\bibitem[\protect\citeauthoryear{Pustilnik et al.}{2021}]{BTA1}
Pustilnik S.A., Egorova E.S., Kniazev A.Y., Perepelitsyna Y.A., Tepliakova A.L., Burenkov A.N., Oparin D.V.,
2021, MNRAS, 507, 494


\bibitem[\protect\citeauthoryear{Pustilnik et al.}{2024}]{SALT2}
Pustilnik S.A., Kniazev A.Y., Tepliakova A.L., Perepelitsyna Y.A., Egorova E.S.,
2024, MNRAS, 527, 11066

\bibitem[\protect\citeauthoryear{Pustilnik et al.}{2024}]{DDO68NR}
  Pustilnik S.A., Perepelitsyna Y.A., Vinokurov A.S., et al.
  2024, Astrophys.Bull., 79, 594 (arXiv:2411.07393)

\bibitem[\protect\citeauthoryear{Pustilnik, Perepelitsyna}{2025}]{DDO68-V1}
  Pustilnik S.A., \& Perepelitsyna Y.A.,
  2025, A\&A, 695, L7

\bibitem[\protect\citeauthoryear{Schlafly, Finkbeiner}{2011}]{SF2011}
  Schlafly E.F., \&  Finkbeiner D.P., 2011, ApJ, 737, article id. 103

\bibitem[\protect\citeauthoryear{Skillman et al.}{2013}]{Skillman13}
Skillman E., Salzer J.J., Berg D.A., et al., 2013, AJ, 146, 3

\bibitem[\protect\citeauthoryear{Stasinska}{2005}]{Stasinska05}
  Stasinska G., 2005, A\&A, 434, 507-520

\bibitem[\protect\citeauthoryear{Telford et al.}{2023}]{Telford23}
Telford O.G., McQuinn K.B.W., Chisholm J., Berg D.A.,
2023, ApJ, 943:65 (17pp)

\bibitem[\protect\citeauthoryear{Telford et al.}{2024}]{Telford24}
Telford O.G., Chisholm J., Sander A.A.C., Ramachandran V., McQuinn K.B.W., Berg D.A.,
2024, ApJ, 974:85 (22pp)

\bibitem[\protect\citeauthoryear{Tramper et al.}{2015}]{Tramper2015}
Tramper F., Straal S.M., Sanyal D.,  et al.
2015, A\&A, 581, A110

\bibitem[\protect\citeauthoryear{Tully et al.}{2009}]{Tully2009}
Tully R.B., et al., 2009, ApJ, 138, 323

\bibitem[\protect\citeauthoryear{Tweed et al.}{2018}]{Tweed18}
Tweed D.P., Mamon G.A., Thuan T.X., Cattaneo A., Dekel A., Menci N., Calura F., Silk J.,
2018,   MNRAS, 477, 1427

\bibitem[\protect\citeauthoryear{Zackrisson et al.}{2001}]{Zackrisson01}
Zackrisson E., Bergvall N., Olofsson K., Siebert A., 2001, A\&A, 375, 814

\end{thebibliography}
\end{document}